\documentclass[10pt,journal,compsoc]{IEEEtran}

\usepackage{graphicx}
\usepackage[utf8]{inputenc} 
\usepackage{url}            
\usepackage{booktabs}       
\usepackage{amsfonts}       
\usepackage{nicefrac}       
\usepackage{microtype}      
\usepackage[inline]{enumitem}
\usepackage[final]{changes}

\usepackage[noend,ruled,linesnumbered]{algorithm2e}
\usepackage{algorithmic}

\usepackage{amsfonts}
\usepackage{lipsum}
\usepackage{graphicx}
\usepackage{amstext}
\usepackage{amsmath}
\usepackage{mathrsfs}
\usepackage{enumitem}
\usepackage{amsthm}
\usepackage{amsmath,amsfonts,amssymb,amsthm,epsfig,epstopdf,url,array}
\usepackage{multirow}
\usepackage{enumitem}
\usepackage{graphicx}
\usepackage[misc]{ifsym}

\usepackage{amsmath,amsfonts,bm}

\usepackage{xspace}

\makeatletter
\DeclareRobustCommand\onedot{\futurelet\@let@token\@onedot}
\def\@onedot{\ifx\@let@token.\else.\null\fi\xspace}

\def\etc{\emph{etc}\onedot}

\makeatother

\def\eqref#1{equation~\ref{#1}}

\def\1{\bm{1}}

\def\vh{{\bm{h}}}

\def\mS{{\bm{S}}}

\DeclareMathAlphabet{\mathsfit}{\encodingdefault}{\sfdefault}{m}{sl}
\SetMathAlphabet{\mathsfit}{bold}{\encodingdefault}{\sfdefault}{bx}{n}

\DeclareMathOperator*{\argmax}{arg\,max}

\usepackage[backend=bibtex,style=ieee,natbib=true,mincitenames=1]{biblatex} 
\setitemize{noitemsep,topsep=0pt,parsep=0pt,partopsep=0pt}

\newtheorem{problem}{Problem}
\theoremstyle{definition}

\usepackage{xspace}
\mathchardef\mhyphen="2D

\newcommand{\eat}[1]{}

\makeatletter
\DeclareRobustCommand\onedot{\futurelet\@let@token\@onedot}
\def\@onedot{\ifx\@let@token.\else.\null\fi\xspace}

\def\etc{\emph{etc}\onedot} 
 
\newcommand{\xhdr}[1]{\noindent{{\bf #1.}}}
\usepackage{soul}
\usepackage{booktabs}
\usepackage{subfig}
 
\ifCLASSINFOpdf
\else
\fi

\begin{document}
%
\title{Revisiting Adversarial Attacks on Graph Neural Networks for Graph Classification}
%
%
%
%
\author{Xin~Wang,~\IEEEmembership{Member,~IEEE},
        Heng~Chang,
        Beini~Xie,
        Tian~Bian,
        Shiji~Zhou,
        Daixin~Wang,
        Zhiqiang~Zhang,
        and~Wenwu~Zhu,~\IEEEmembership{Fellow,~IEEE}
\IEEEcompsocitemizethanks{

\IEEEcompsocthanksitem X. Wang and W. Zhu are with the Department of Computer Science and Technology, BNRist, Tsinghua University, Beijing 100084, China.
E-mail: \{xin\_wang, wwzhu\}@tsinghua.edu.cn.
\IEEEcompsocthanksitem B. Xie, H. Chang, S. Zhou are with the Tsinghua-Berkeley Shenzhen Institute, Tsinghua University, Shenzhen 518055, China.
E-mail: \{xbn20, changh17, zsj17\}@mails.tsinghua.edu.cn
\IEEEcompsocthanksitem T. Bian is with the System Engineering and System Management Department, Chinese University of Hong Kong, Hong Kong 999077, China.
E-mail: tianbian@link.cuhk.edu.hk
\IEEEcompsocthanksitem D. Wang and Z. Zhang are with the Ant Group, Hangzhou 310063, China.
E-mail: \{daixin.wdx, lingyao.zzq\}@antgroup.com
\protect
}
\thanks{This work was supported by the National Key Research and Development Program of China No. 2020AAA0107801, National Natural Science Foundation of China (No. 62222209, 62250008, 62102222), Beijing National Research Center for Information Science and Technology under Grant No. BNR2023RC01003, BNR2023TD03006, 
and Beijing Key Lab of Networked Multimedia. (Corresponding authors: Xin Wang and Wenwu Zhu.)}

}

%
%

\markboth{IEEE TRANSACTIONS ON KNOWLEDGE AND DATA ENGINEERING}%
{\MakeLowercase{Xin Wang~\textit{et al.}}: Revisiting Adversarial Attacks on Graph Neural Networks for Graph Classification}
%



\IEEEtitleabstractindextext{%
\begin{abstract}
Graph neural networks (GNNs) have achieved tremendous success in the task of graph classification and its diverse downstream real-world applications.
Despite the huge success in learning graph representations, current GNN models have demonstrated their vulnerability to potentially existent adversarial examples on graph-structured data.
Existing approaches are either limited to structure attacks or restricted to local information, 
urging for the design of a more general attack framework on graph classification, which faces significant challenges due to the complexity of generating \textit{local-node-level} adversarial examples using the \textit{global-graph-level} information. 
To address this "global-to-local" attack challenge, we present a novel and general framework \textit{CAMA} to generate adversarial examples via manipulating graph structure and node features. 
Specifically, we make use of Graph Class Activation Mapping and its variant to produce node-level importance corresponding to the graph classification task. 
Then through a heuristic design of algorithms, we can perform both feature and structure attacks under unnoticeable perturbation budgets with the help of both node-level and subgraph-level importance. 
Experiments towards attacking four state-of-the-art graph classification models on six real-world benchmarks verify the flexibility and effectiveness of our framework.

\end{abstract}

}

\maketitle

\IEEEdisplaynontitleabstractindextext

%
\IEEEpeerreviewmaketitle

\IEEEraisesectionheading{\section{Introduction}\label{sec:introduction}}

\IEEEPARstart{G}{raph} structured data is ubiquitous for capturing relations and interactions at the level of node classification~\cite{ICLR2017SemiGCN}, edge prediction~\cite{zhang2018link}, and graph classification~\cite{Gilmer2017Neural}. 
Among them, graph classification plays a vital role in a wide range of domains~\cite{zhang2018deep}.

For instance, in social network analysis, the fake news detection problem can be regarded as a binary graph classification task over Twitter's news propagation networks~\cite{monti2019fake}. As a powerful tool with the expressive capability of deep learning on graph data, the family of Graph Neural Networks (GNNs) has gained tremendous popularity over the past few years in graph classification and its downstream real-world applications~\cite{gomez2017dynamics,kim2019hats,magelinski2020graph,li2018deep, li2020weakly, li2021ctnet}.

\begin{figure}[t]
    \centering
    \includegraphics[width=0.45\textwidth]{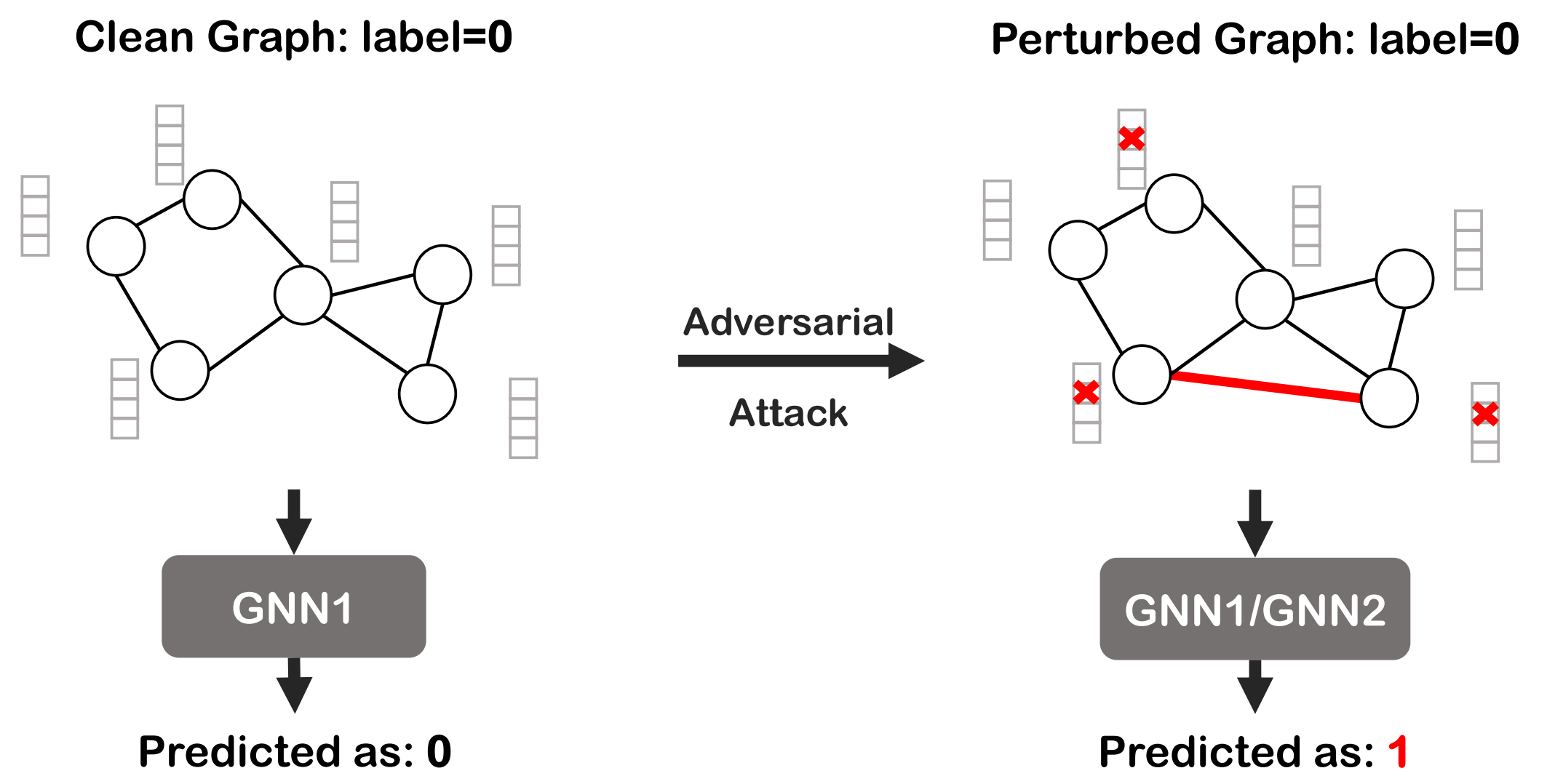}
    \caption{\textbf{Adversarial attack on graph classification}. Given a cleaned graph, we can manipulate node features and edges to generate a poisoned graph to fool the victim GNN.}
    \label{fig: camatk}
\end{figure}

Despite the powerful ability of GNNs in learning graph representations, their vulnerability to potentially existent adversarial examples on graph-structured data has been revealed recently~\cite{jin2020graph}.
Therefore, the lack of robustness within GNNs may be exploited by fraudsters or spammers, potentially provoking dissent on their applications in security-critical domains.
For example, deliberately modifying personal identity information without authorization will result in credit card fraud~\cite{BHATTACHARYYA2011602}.  
Similar to the utilization of graph-structured data, adversarial attacks on graphs can also be broadly categorized into node level and graph level, in terms of the type of different tasks. 
On the one hand, studies towards node-level adversarial attacks are quite comprehensive from various perspectives~\cite{xu2019topology,dai2018adversarial,chang2020restricted,wu2019adversarial,bojchevski2019adversarial}.
On the other hand, in contrast to the remarkable and relatively mature frameworks for adversarial attacks on node-level tasks, systematic research regarding a general attacking framework for adversarial attacks on graph classification tasks is largely unexplored regardless of the vast importance.

\begin{figure*}[t]
    \centering
\includegraphics[width=0.9\textwidth]{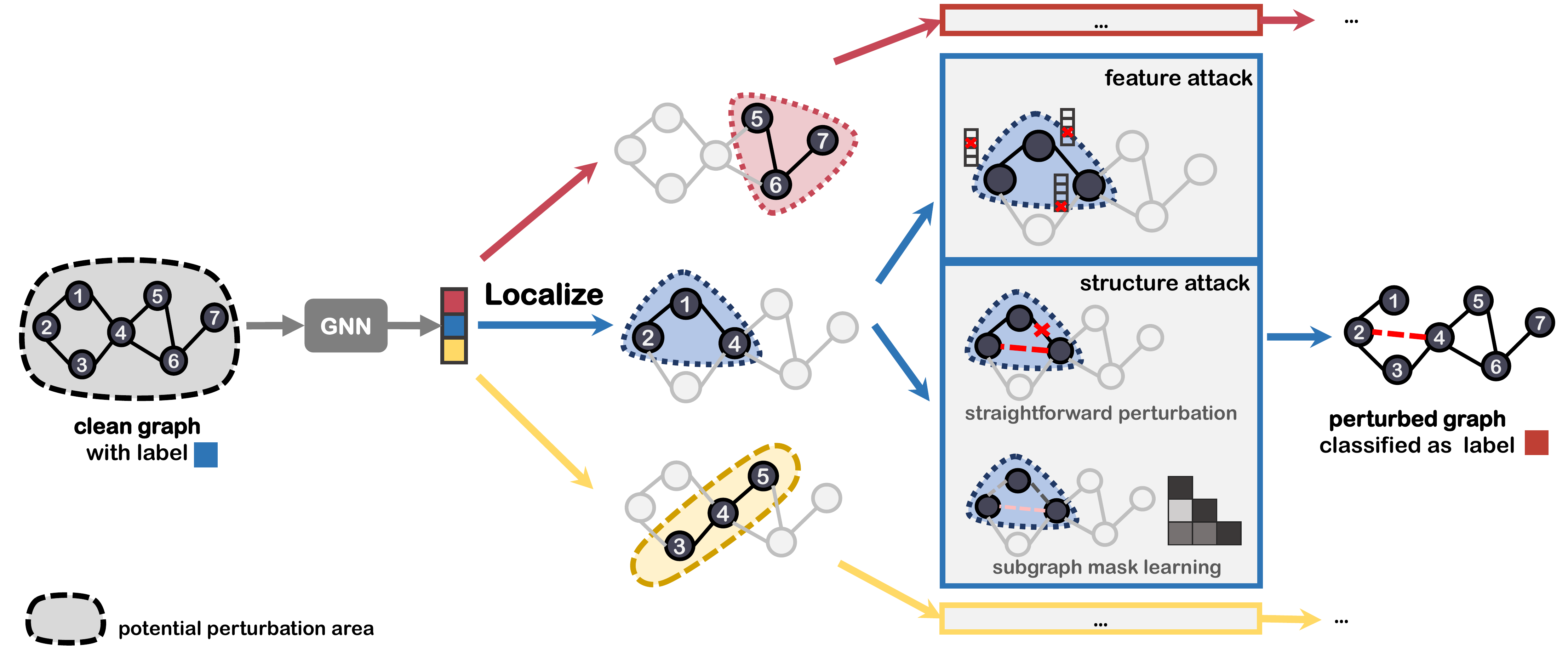}
    \caption{
    \textbf{An example of \textit{CAMA} for a three-classes graph classification task.} 
    In the CAMA framework, we first \textbf{localize} potential perturbations to top-ranked important nodes. Then, we generate corresponding perturbed graphs until obtaining a successful feature/structure attack.}
    \label{fig: CAMA example}
\end{figure*}

Compared with perturbations for node-level classification, migrating these adversarial examples to graph-level tasks is a non-trivial problem, since they have different goals for optimization from the local to global scale.
We denote this problem as the "global-to-local attack challenge" which is illustrated in detail in Section~\ref{sec: challenge}.
The desired attack framework for graph classification should be general to conduct both feature and structure attacks. Moreover, the effective attack on one graph classification model is expected to be able to be successfully transferred to other graph classifiers (an illustration is shown in Figure~\ref{fig: camatk}).
Most importantly, a successful attacker is expected to perform unnoticeable attacks and effectively localize important nodes/edges to perturb via the global-level classification information.
In a nutshell, the research on graph classification adversarial attacks still faces \textbf{three main challenges}:
\begin{itemize}
\item Given that graph classification tasks depend on efficiently learning global graph representation from local node embeddings via pooling functions, it is complex to exploit information of global-graph-level classification to generate local-node-level adversarial examples;
\item Most existing approaches can only attack graph structure. However, node features might contain more fruitful information. For example, personal identity information and loan history apparently matter more in credit models and fraud detection. 
Therefore in a more realistic and practical condition, we need a general attack framework that is able to manipulate both node features and graph structure;
\item 
Current attack methods for graph classification using gradient information only consider the training of target models and fail to reflect the information from the global graph structure, which might easily result in the generated adversarial edges being trapped around a single node 
or concentrating on high-degree nodes as we observe from experiments. 
\end{itemize}

To tackle these challenges, we propose a novel hierarchical framework, namely \textbf{\textit{CAMA}}, to bridge the gap between local-node-level and global-graph-level information. We migrate the idea from \textbf{C}lass \textbf{A}ctivation \textbf{M}apping (CAM)~\cite{pope2019explainability} to conduct powerful adversarial \textbf{A}ttack towards graph classification tasks. This unified solution sheds light on the problem of quantifying the contribution of local node information to global representation in attacking graph classification tasks.
An example of \textit{CAMA} is shown in Figure~\ref{fig: CAMA example}.
To summarize, our work makes the following \textbf{main contributions},
\begin{itemize}
\item \textbf{Framework:} We propose the novel \textit{CAMA} framework for adversarial attacks on graph classification.
Our attack approach fills the gap in generating local perturbation examples from global graph classification as well as performs attacks unnoticeably. 
Given the simplicity and effectiveness, \textit{CAMA} can serve as a strong benchmark for future works in this branch.
\item \textbf{Algorithm:} We heuristically design novel algorithms to select target nodes in a graph by graph class activation mapping and its variant, then generate adversarial examples in the level of both structure and feature.
\item \textbf{Experiment:} We show that our method is able to deteriorate graph classification performance by a significant margin on various benchmarks via targeting multiple state-of-the-art GNNs. Further, except for white-box attacks, we also test the transferability of our attack method under the black-box setting for evasion attacks.
\end{itemize}

\section{Related Work}

\xhdr{GNNs on graph classification}
GNNs have proliferated in recent years for tasks like node classification, link prediction, graph classification, and graph generation.
GNNs often stack multiple graph convolutions followed by a readout operation to aggregate nodes' information to a graph-level representation when dealing with graph classification tasks.

Various graph convolution layers and graph pooling operations are proposed to learn both nodes and graph representation better~\cite{chang2020spectral,guan2021autogl}.
One of the most popular GNNs is Graph Convolutional Networks (GCN) \cite{ICLR2017SemiGCN} which is inspired by the first-order approximation of Chebyshev polynomials in ChebNet.  It updates the node representation by taking an average representation of their one-hop neighbors. GCN has excellent results in the semi-supervised node classification tasks.
Graph Isomorphism Network (GIN) \cite{xu2019how} uses sum aggregation and multi-layer perceptrons instead of one single activation function. It has excellent discriminative power equal to that of the WL test.
Facing the finite nature of recurrent GNNs, Implicit Graph Neural Network (IGNN) \cite{gu2020implicit} is able to capture long-range dependencies and performs well in both graph classification and node classification on heterogeneous networks. Its framework ensures well-posedness based on Perron-Frobenius's theory.

Except for novel graph convolution operations, diverse pooling strategies affect graph tasks differently. Direct pooling methods like simple node pooling (node-wise mean-pooling, sum-pooling, and max-pooling) directly generate graph-level representation based on node representations \cite{zhou2020graph}. In contrast, hierarchical graph pooling exploits the hierarchical graph structure. 
DiffPool \cite{ying2018hierarchical} proposes a differentiable hierarchical clustering algorithm to learn representations of the new coarsened graph by training a soft cluster assign matrix in each layer. 
Based on the graph Fourier transform, EigenPooling \cite{ma2019graph} jointly uses node features and local structure. 
The graph pooling layer (gPool) \cite{gao2019graphunets} conducts down-sampling on graph data by selecting top-k nodes from calculated projection value. Inversely, the graph unpooling layer (gUnpool) does up-sampling to restore graphs to their original structure. Inspired by U-Net in computer vision, graph U-Nets (g-U-Nets) \cite{gao2019graphunets} is proposed using gPool and gUnpooling operations. g-U-Nets can encode and decode high-level features for network embedding. 

In this paper, we use GCN, GIN, and IGNN as representatives of general graph classification neural networks and use g-U-Nets to represent hierarchical graph classification models.

\xhdr{Adversarial attacks on graph classification}
GNNs have shown their vulnerability under adversarial attacks~\cite{jin2020adversarial}. Most recent works aim to attack models on node classification tasks~\cite{zugner2018adversarial,zugner_adversarial_2019,chang2020restricted,ma2020towards}. Despite their fruitful progress, these methods can only perform attacks on node-level tasks.

For graph-level tasks, based on reinforcement learning, \textit{RL-S2V} \cite{dai2018adversarial} flips edges by selecting two endpoints under black-box attack.
\textit{ReWatt}~\cite{10.1145/3447548.3467416} proposes to perform unnoticeable attacks via rewiring operation and utilizes a similar reinforcement learning strategy as \textit{RL-S2V}.
\textit{Grabnel}~\cite{wan2021adversarial} exploits bayesian optimization to conduct adversarial attacks targeting graph classification models.
Under the white-box setting, \textit{GradArgmax}~\cite{dai2018adversarial} exploits gradients over the adjacency matrix of classification loss and flips edges with the largest absolute gradient.
\textit{Projective ranking}~\cite{10.1145/3459637.3482161} generates adversarial examples by ranking potential edge perturbation masks through encoding node features and projecting selected edge masks.

Nevertheless, the above methods cannot perturb node features.
Further, \cite{tang2020adversarial} proposes an attacking strategy on hierarchical graph pooling neural networks. 
However, they overlook the importance of direct pooling, like simple node pooling. Thus, this approach loses its strength when the graph classification model is unknown.
A novel generic attack framework \textit{GraphAttacker} is recently proposed by~\cite{chen2021graphattacker}, which could attack multiple tasks.
But the time complexity serves as its main concern due to the process of training the GAN-based model.

Considering all of these, adversarial attacks on graph classification are not been fully explored by previous studies. To mitigate this gap, our proposed general framework could flexibly perform structure attacks and feature attacks. Aside from the white-box attack, we also analyze the transferability of our method under black-box attacks.

\xhdr{CAM on graphs}
Class Activation Mapping (CAM) localizes image-level classification into pixel-level image areas by using global average pooling (GAP) in convolutional neural networks in computer vision when it was firstly proposed~\cite{zhou2016learning}. CAM has a strong discriminative localization ability in the explanation of image classification. For example, it can localize the toothbrush region in a picture classified as brushing teeth. 
Compared with the blossom of grand application in computer vision, the utilization of CAM on graph-structured data (Graph CAM) is quite rare with only being applied to the explainability in GNNs~\cite{pope2019explainability,yuan2020explainability}.
Given a graph classification task, Graph CAM can localize the most influential nodes for classification, which then helps us better understand GNNs.
Grad Class Activation Mapping on graphs (Graph Grad CAM)~\cite{pope2019explainability} extends CAM on graphs by loosening architecture restrictions and using gradients of hidden layers as projection weights. 
In this work, we first integrate the localization ability of Graph CAM with the awareness of adversarial attacks on the graph classification tasks. We will undoubtedly increase the scope of research on Graph CAM.

\section{Preliminaries}
\subsection{Notations}
Given a set of graphs $\mathcal{G}=\{G_i\}_{i=1}^N$, where $|\mathcal{G}|=N$, we consider graph classification on $\mathcal{G}$. Each graph  $G_i = (\bm{A}_i, \bm{X}_i)$ has $n_i$ nodes, where $\bm{A}_i \in \{ 0,1\}^{n_i \times n_i}$ is the adjacency matrix and $\bm{X}_i \in \mathbb{R}^{n_i \times D} $ is the node feature matrix with dimension $D$. Each $G_i$ is assigned with a label $c_i \in \mathcal{C}=\{ 1,2,...,C\}$, where $C$ is the total number of classes.

\subsection{Graph Classification}
Graph classification aims to predict the labels of unlabeled graphs. With paired graphs and labels $\{ G_i, c_i\}_{i=\{1,\cdots, N \}}$, its goal is to learn a mapping function $f: \mathcal{G} \rightarrow \mathcal{C}$. 
We simplify graph classification model architecture and consider only one fully connected layer. Given a graph  $G_i=(\bm{A}_i,\bm{X}_i)$ with $n_i$ nodes, a standard procedure for graph classification with direct pooling can be formulated as:
\begin{align}
\bm{h}^{(0)}_i &= \bm{X}_i,  \; \;
\bm{h}^{(l)}_i = f_{conv}(\bm{h}^{(l-1)}_i;\bm{\Theta}^l), l=1,2,...,L    \label{nodeembed} \\
\bm{h}_i &= \operatorname{pooling}(\bm{h}^{(L)}_i),  \; \;
\bm{z}_i = \bm{W} \bm{h}_i+\bm{b},    \label{weight}
\end{align}
where $\bm{h}_i^{(l)} \in \mathbb{R}^{n_i \times D_l}$ denotes the hidden node embedding in the $l$-th graph convolution $f_{conv}$,  and $\bm{\Theta}^l$ is the corresponding parameter matrix. $\bm{h}_i\in \mathbb{R}^{D_L}$ is the graph embedding of $G_i$ after pooling of final node embedding $\bm{h}_i^{(L)} \in \mathbb{R}^{n_i \times D_L}$. $\bm{W} \in \mathbb{R}^{C \times D_L}$ and $\bm{b} \in \mathbb{R}^{C}$ are parameters in the output fully connected layer, and $L$ is the number of graph convolution.

The objective function for graph classification can be further formulated as:
\begin{equation*}
\operatorname{min}_{\bm{\Theta}}\mathcal{L} _{\bm{\Theta}}(\mathcal{G})  = \sum_{i=1}^N
l(f_{\bm{\Theta}}(G_i),c_i),
\end{equation*}
where $l(\cdot,\cdot)$ is a loss function such as the cross-entropy.

\subsection{Adversarial Attacks towards GNNs}

The problem of adversarial attacks on graph classification is to misclassify graph labels, which is formulated as follows:

\begin{problem}
Given paired data of graphs and their labels $\{ G_i, c_i\}_{i=1}^N$, the goal of an attacker is to minimize the attack objective function $\mathcal{L}_{atk}$:

\begin{equation*}
\operatorname{argmin}_{\mathcal{G}'}\mathcal{L}_{atk}(\mathcal{G}') = \sum_{i=1}^N l_{atk}(f_{\bm{\Theta}}(G_i'), c_i),     
\end{equation*}
where $l_{atk}$ is the attack loss function, and $G_i'$ denotes the perturbed version of $G_i$.
\end{problem}

We could define $l_{atk}=-l$ where $l$ is set as the cross-entropy loss for graph classification. We can also define $l_{atk}$ as the other attack loss like the CW-loss~\cite{zugner2018adversarial}.

In the real world, the attacker usually only unnoticeably attacks within perturbation budget $\Delta$ for each graph $G_i$. Thus, the domain of modified graphs is constrained as :
\begin{equation*}
\parallel \bm{A}_i'-\bm{A}_i\parallel_0 + \parallel \bm{X}_i'-\bm{X}_i\parallel _0\leq \Delta, 
\end{equation*}
where $\bm{A}'_i$ and $\bm{X}'_i$ is the perturbed adjacency matrix and node feature matrix for graph $G'_i$.
In the following sections, we omit the subscript $i$ for graph $G_i$ for simplicity. 

Adversarial attacks have various taxonomies from the perspectives of perturbation type (feature attack and structure attack), attacker's knowledge (white-box attack and black-box attack), and the stage where attacks happen (evasion attack and poisoning attack).
A desired general framework should be able to deliberate most situations mentioned above, which is also the aim of this work.

\section{Methodology}
In this section, we start by introducing the global-to-local attack challenge.
In order to tackle this challenge, we propose to first localize potential perturbations to top-important nodes and then perform attacks targeting these important nodes. The whole attack process is decomposed into two steps: 1) \textit{node importance estimation}, and 2) \textit{adversarial example generation}. 
In this way, we are able to first transfer the focus from classification on graphs to the contribution of each node and design perturbations locally afterward.

\subsection{Global-to-Local Attack Challenge}
\label{sec: challenge}

Generating adversarial examples toward graph classification is intrinsically a global-to-local problem. The goal of attackers is to fool the GNNs from correct predictions on graph-level labels.
However, the adversarial attacks must be localized to node and edge levels.
The global-to-local problem is non-trivial to solve since graph-level predictions and node-level attacks are implicitly bridged via the pooling functions in GNNs.
As empirical evidence, we observe that existing methods either give more attention to high-degree nodes or easily get trapped around one single node, which makes the adversarial examples noticeable and then undesirable. 

\begin{table}[htbp]
\caption{Average degree of the selected nodes in diverse feature attack methods.}
\label{tab: avg degree}
\centering
\resizebox{0.99\columnwidth}{!}{
\begin{tabular}{lccccccc}
\toprule
Dataset & \textit{Random} & \textit{Degree} & \textit{Betweenness} & \textit{RWCS} & \textit{GC\_RWCS} \\
\midrule
MUTAG & 2.23 & 3.00 & 2.99 & 2.99 & 2.72\\
COX2 & 2.00 & 3.24 & 3.01 & 3.18 & 2.49 \\
\bottomrule
\end{tabular}
}
\end{table}

Table~\ref{tab: avg degree} shows the comparison of the average node degree of nodes selected by different attack methods.
Compared with \textit{Random} which selected nodes to attack randomly, the other attack methods tend to select nodes with higher degrees, which makes the attack process more noticeable.

\begin{figure}[t]
\centering
\includegraphics[width=0.99\columnwidth]{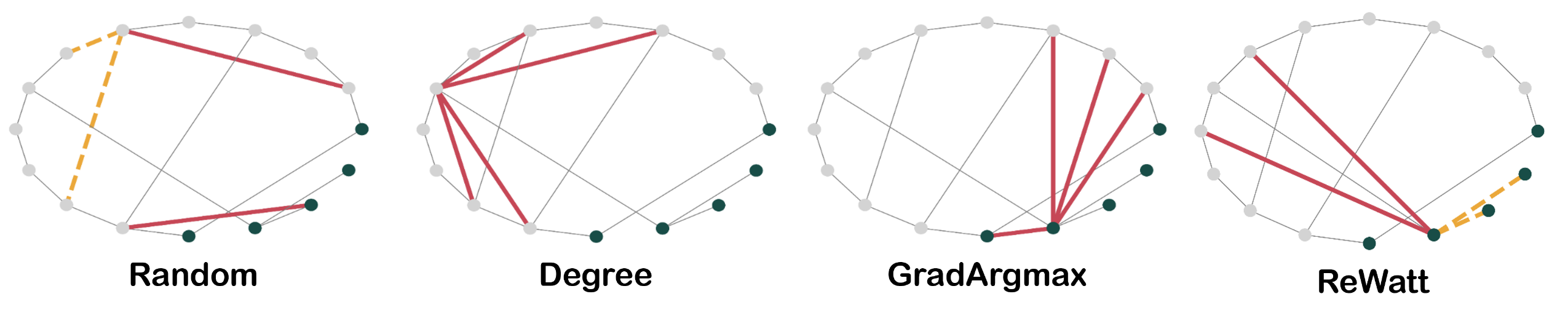}
\caption{\textbf{An example of structure attack on MUTAG dataset with edge attack proportion=$20\%$.}
Added edges are shown in red lines and deleted edges are shown in orange dashed lines.}
\label{fig: comparison}
\vspace{-5mm}
\end{figure}

In addition, we visualize selected important nodes and generated structural perturbations from various baselines in Figure~\ref{fig: comparison}.
For example, for \textit{Degree}, the perturbed edges are selected based on node degree. This may result in irrelevant perturbation between nodes and edges. Meanwhile, we can observe that the adversarial edges produced by \textit{GradArgmax} and \textit{ReWatt} are trapped near one single node, which further implies the deficiency of these two methods that extract merely local information.

As a result, this paper quantifies the contribution of nodes at the local level to graph classification tasks at the global level and reversely conducts effective adversarial attacks at the local level to destroy the global level classification performance. 
Our methods could better utilize the graph-level classification information to localize to nodes/edges not only important to the classification but also irrelevant to several super influential nodes.

\subsection{Node Importance Estimation}
The node importance estimation helps to localize potential perturbations into high-important nodes in a graph. 
Specifically, after finishing model training, we determine the contribution of nodes from the local level to graph classification in a way inspired by Graph CAM and its variant~\cite{pope2019explainability}.

\xhdr{Graph CAM} As a useful method that provides explainability for graph classification, Graph CAM has been well studied. Since the weight matrix of the output fully connected layer can represent the importance of the features of each dimension for graph classification, Graph CAM builds a heat-map matrix by projecting back the weight matrix to the node representation in the final graph convolution layer to indicate the importance of each node for graph classification. This heat-map matrix is calculated as:
\begin{equation}
\label{heat map}
\bm{L}_{CAM} = \operatorname{ReLU}(\bm{h}^{(L)} \bm{W}^T),    
\end{equation}
where $\bm{W}\in \mathbb{R}^{C \times D_L}$ is the same weight matrix as in Eq.~\eqref{weight}, and $\bm{h}^{(L)} \in \mathbb{R}^{n \times D_L}$ denotes node representation in the final graph convolution layer for one graph as shown in Eq.~\eqref{nodeembed}. The $k$-th element in $c$-th row of $\bm{W}$ indicates the importance of feature $k$ for predicting label $c$.

A variant of Graph CAM, Graph Grad CAM, uses gradients with respect to each hidden convolutional layer and each class. Then, the calculation of gradients $\bm{\alpha}^{l} \in \mathbb{R}^{D^l \times {C}}$ replaces weights $\bm{W}^T$ in CAM to construct the heat-map matrix for each layer. At last, by taking the average over the heat-map matrix of all graph convolution layers, the heat-map matrix is calculated as:
\begin{align}
\bm{\alpha}^{l} = \frac{1}{n} \sum_v \frac{\partial \bm{z}^T}{\partial \bm{h}_v^{(l)} }, \; \;
\bm{L}_{Grad-CAM} = \frac{1}{L} \sum_l \operatorname{ReLU}(\bm{h}^{(l)}\bm{\alpha}^{l}), \notag
\end{align}
where $\bm{z} \in \mathbb{R}^{C}$ is the prediction logits, $\bm{h}_v^{(l)} \in \mathbb{R}^{D^l}$ is the hidden embedding for node $v$ in the $l$-th graph convolution layer. The $i$-th entry in $c$-th column of $\bm{L}_{CAM}$ indicates the relative importance for node $i$ resulting from classifying $G_i$ into class $c$.  

Though having great explainability, directly using Graph CAM still has two limitations.
Firstly, the number of fully connected layers is fixed to one due to the restriction on matrix multiplication in Graph CAM.
Secondly, the hidden size must be kept the same for all hidden convolutional layers for Graph Grad CAM.
As we will show in experiments, these architecture restrictions do not deteriorate classification performance on clean graphs. Also, they do not hinder the transferability of our proposed attack methods.

\xhdr{Ranked CAM Matrix}
We calculate the ranked CAM matrix based on the CAM heat-map matrix. The whole process is summarized in Algorithm~\ref{alg:rank}.
After getting the CAM heat-map matrix, 
We first rank each column in descending order and get the corresponding nodes ranking matrix $\bm{U}_{CAM}^{orig} \in \mathbb{R}^{n \times {C}}$ in line 1. This implies the class-specific view of node importance ranking. Then, we exploit $\bm{U}_{CAM}^{orig}$ to calculate a global-level nodes ranking vector $\bm{u}_{global}\in \mathbb{R}^{n}$ in line 2. 
Specifically, we go through each row in $\bm{U}_{CAM}^{orig}$ and use the 
highest ranking among all columns for each node until all nodes are included in $\bm{u}_{global}$.
Finally, we concatenate these two ranking sources of nodes to get the final ranked CAM matrix $\bm{U}_{CAM}\in \mathbb{R}^{n \times (C+1)}$.

\begin{algorithm}[htb]
\caption{Generating ranked CAM matrix.}
\label{alg:rank}
\KwIn{Heat-map matrix $\bm{L}_{CAM}$.} 
\KwOut{Ranked CAM matrix $\bm{U}_{CAM}$.}
\LinesNumbered
$ \bm{U}_{CAM}^{orig} \gets \operatorname{column\_rank}(\bm{L}_{CAM})$;

$\bm{u}_{global} \gets \operatorname{global\_rank}(\bm{L}_{CAM})$;

$ \bm{U}_{CAM} \gets  \operatorname {concatenate}([\bm{U}_{CAM}^{orig},\bm{u}_{global}])$;
    
Return $ \bm{U}_{CAM}$;

\end{algorithm}

Because the CAM heat-map matrix can precisely demonstrate the importance of each node for graph classification tasks, after the ranking operation on CAM heat-map, each column in $\bm{U}_{CAM}$ indicates one type of view for nodes' importance ranking. We could identify the most influential nodes for the whole graph classification process through different views of the ranked CAM matrix and generate adversarial examples accordingly. Since the adversarial attack depends on Graph CAM, we name our framework as \textbf{CAM} based \textbf{A}ttack and its variant \textbf{\textit{CAMA-Grad}} when using Graph Grad CAM.

\subsection{Adversarial Example Generation}
With access to the ranked CAM matrix $\bm{U}_{CAM}$, we call each column of $\bm{U}_{CAM}$ as the ranked CAM vector, denoted as $\bm{U}^c, c = 1,...,C+1$ . How do we generate adversarial examples with a series of ranked CAM vectors? Here, we heuristically propose two attack algorithms towards \textit{\textbf{CAMA}} (for feature attack and structure attack) and  \textit{\textbf{CAMA-subgraph}} (for structure attack only). 
For \textit{CAMA}, in the overall adversarial perturbation, we repeat using our algorithms for each column of the ranked CAM matrix $\bm{U}^c$ until a successful attack.
For \textit{CAMA-subgraph}, we only need the column of the predicted label in the ranked CAM matrix to select the candidate perturbations.
Both two algorithms have their grad version
\textit{CAMA-Grad} and \textit{CAMA-subgraph-Grad}. The difference between algorithms and their grad version lies only in calculating the CAM heat-map matrix.

\subsubsection{Feature Attack \label{Feature Attack}}
For feature perturbations, we set both global-level and local-level perturbation budgets. In global-level budgets, we assume only a few nodes of one particular graph are available. These nodes are called target nodes. In local-level budgets, we constrain the number of features to be adjusted.

Given the limitation of modified node amount $r$, target nodes are selected by the first $r$ nodes in the ranked CAM vector $\bm{U}^c$. A small constant noise $\epsilon$ is added to each feature of target nodes for perturbation, while $\epsilon$ relies on the attacker's knowledge of node features. 
Specifically, given the information of the training process, the number of adjusted features $K$ and adjusted magnitude $\lambda$, noise $\epsilon_{j}$ added for the $j$-th feature could be calculated following \cite{ma2020towards} as: 
\begin{equation*}
\epsilon_{j}=\left\{\begin{array}{ll}
\lambda \cdot \operatorname{sign}  
\left(\sum_{i=1}^{n} \frac{\partial l(f_{\theta}(G),c)}{\partial X_{i j}}\right), \\
\ \ \ \text{if} j \in  \operatorname{argtop-K} \left(\left[\left|\sum_{i=1}^{n} \frac{\partial l(f_{\theta}(G),c)}{\partial X_{i l}}\right|\right]_{l=1,2, \ldots, D}\right) \\
0, \text { otherwise. }
\end{array}\right.
\end{equation*}
We replace Carlili-Wagner loss in \cite{ma2020towards} with cross-entropy loss. The overall number of perturbations is $rK \leq \Delta $.
We summarize the process of \textit{CAMA} for generating feature perturbations in Algorithm~\ref{alg:feature}.

\begin{algorithm}[htb]
\caption{\textit{CAMA} for feature perturbations.} 
\label{alg:feature}
\KwIn{Graph $G = (\bm{A}, \bm{X})$ with $n$ nodes; number of nodes limit $r$; ranked nodes vector  $\bm{U}^c$; feature noise $\epsilon_j$, where $j=1,2,...,D$.}   
\KwOut{Modified feature matrix $\bm{X}$.}
Initialize modified feature matrix $\bm{X}'\gets \bm{X}$;
$C_{nodes} \gets  \bm{U}^c[:r]$;

\For{$u$ in $C_{nodes}$}{
    $\bm{X}_j'[u] \gets\bm{X}_j[u]+\epsilon_j, \quad j=1,2,...,D$
}

\textbf{return} $\bm{X}'$;
\end{algorithm}

\subsubsection{Structure Attack}
Structure attack is more comprehensive compared with feature attack, considering the complexity of connectivity in graphs. To this end, 
we specially design two structure attack algorithms: \textit{CAMA} and \textit{CAMA-subgraph}, with the help of the ranked CAM matrix. \textit{CAMA} is an efficient algorithm that performs attacks via simply flipping edges among top-ranked vital nodes in the ranked CAM matrix.
\textit{CAMA-subgraph} then takes a step further to attack by learning a subgraph mask to select edges for perturbation.

\xhdr{\textit{CAMA}: straightforwardly flipping edges among most important nodes}
To generate structure perturbations,  we assume edges among nodes of higher activation importance are more influential in graph classification tasks and intuitively flip edges among them.
With the known ranked nodes' influence on graph classification, we flip edges among nodes that have a higher ranking.  
Furthermore, we exploit node similarity to enhance attack ability aside from the information from the graph structure. The similarity score is calculated as follows.

Given a learned embedding of node $\bm{h}^{emb}$, similarity $\bm{S}$ between nodes $u$ and $v$ is calculated with cosine distance: 
\begin{equation*}
    \bm{S}[u,v]=\mS[v,u]=\cos({\bm{h}}^{emb}_u,{\bm{h}}^{emb}_v).
\end{equation*}
We constrain the operation of adding/deleting edges within the similarity constraint.
Under the graph homophily assumption and with the  calculated similarity matrix $\mS$, we choose to add edges between low-similarity node pairs and delete edges between high-similarity node pairs: 
$$
\left\{
\begin{aligned}
\bm{A}'[u,v] -\bm{A}[u,v]&=1, \bm{A}[u,v]=0   \  \operatorname{and} \ \bm{S}[u,v] \leq s_1;\\
\bm{A}'[u,v] -\bm{A}[u,v]&=-1, \bm{A}[u,v]=1 \ \operatorname{and} \ \bm{S}[u,v] \geq s_2. 
\end{aligned}
\right.
$$

Our attacking strategy is in heuristic way by increasing the ranking number each time, iteratively finding candidate pairs of nodes, and flipping edges between new target nodes and old ones within perturbation budget and similarity restriction.
In each iteration, we increase ranking number $i$ by one and add a new node $u_i$, which ranked $i$-th in vector $\bm{U}^c$, into target nodes set $C_{nodes}$. In the end, we flip edges between new target nodes and old ones within $\Delta$.
The overall procedure for structure perturbations is summarized in Algorithm~\ref{alg:structure}.

\begin{algorithm}[htb]
\caption{\textit{CAMA} for structure perturbations.}
\label{alg:structure}
\KwIn{Graph $G = (\bm{A}, \bm{X})$ with $n$ nodes; modification budget $\Delta$; Similarity matrix $\bm{S}$; similarity restriction parameter $s_1$, $s_2$; ranked nodes vector  $\bm{U}^c$.} 
\KwOut{Modified adjacency matrix $\bm{A}'$.}
\LinesNumbered
Initialize remaining perturbation number $n_{perturbs}\gets \Delta$, modified adjacency matrix $\bm{A}'\gets \bm{A}$, target nodes set $C_{nodes}= \bm{U}^c[0]$, and current rank index $i=1$.

\While {($i \leq n$) and ($n_{perturbs} > 0$)} {
    $ u_i \gets\bm{U}^c[i]$;\\
    \For {$v$ in $C_{nodes}$}
    {
        \If{$\operatorname{similarity\_constraint}((u_i, v); \bm{S}, s_1, s_2)$}
        {
            $\bm{A}'[u_i,v]\gets 1-\bm{A}[u_i,v]$,\\$n_{perturbs}\gets n_{perturbs}-1$;\\
            \If{$n_{perturbs} == 0$}{break;\\}
        }
    }
    $C_{nodes}\gets [ C_{nodes}, u_i]$; \\
    $i \gets i+1$;
}
Return $\bm{A}'$;

\end{algorithm}

\xhdr{\textit{CAMA-subgraph}: attack with subgraph mask learning}
In order to further exploit the local information from a subgraph perspective, we propose an end-to-end adversarial structure attack model with subgraph mask learning.

For each graph $G$, we obtain a subgraph $G_{sub}$ by keeping $p\%$ top ranked nodes $\mathcal{V}_{sub}, |\mathcal{V}_{sub}| = \lfloor p\%|\mathcal{V}| \rfloor$ in the nodes rank vector with view of predicted label $c$ (the $c$-th column $\bm{U^c}$ in the ranked CAM matrix). 
Then, we limit potential edge perturbations $\bm{M}=\mathcal{V}_{sub} \times \mathcal{V}_{sub}$ within the subgraph. With the edge perturbation candidates $\{m_{uv}| u,v \in \mathcal{V}_{sub}, \sum_{uv} m_{uv} \leq \Delta\}$, the adversarial examples are calculated as follows:

\begin{equation}
    c_{uv} = 1-2a_{uv}
\end{equation}
\begin{equation}
a'_{uv}=
    \left\{
    \begin{array}{ll}
    a_{uv}+ c_{uv}\sigma(m_{uv}), \;& u \in  \mathcal{V}_{sub}, v\in \mathcal{V}_{sub} \\
    a_{uv}, \; & others,
    \end{array}
    \right.
\end{equation}
where $\sigma(\cdot)$ is the sigmoid function to map mask values into zero and one. The larger value of $m_{uv}$, the more attack importance to perturb edge $a_{uv}$.

Given a trained victim model $f_{\bm{\Theta}}$, we minimize the attack loss $l_{atk}$ for each graph with the victim model's parameters unchanged to learn the subgraph mask $m_{uv}$:
\begin{align}
    \min l_{atk} = l_{cw}+\lambda_{ent}l_{ent},
\end{align}
where $l_{cw}$ denotes for CW-loss, and $l_{ent}$ represents the mean entropy of each element $m_{uv}$.
$l_{cw}$ aims to achieve a successful attack~\cite{zugner2018adversarial} while $l_{ent}$ encourages the masking value of $\sigma(m_{uv})$ to be binary~\cite{ying2019gnnexplainer}.
Hyper-parameter $\lambda_{ent}$ balances the influence of $l_{cw}$ and $l_{ent}$ in the total loss function.

Specifically, given the ground truth label $c_{yt}$ of the graph, the detailed designs of $l_{cw}$, $l_{ent}$ are:
\begin{align}
    l_{cw} = \max(z_{c_{yt}}-\max_{c'\neq c_{yt}} z_{c'}, 0),
\end{align}
\begin{align}
    \begin{aligned}
     l_{ent} = & -\frac{1}{|\bm{M}|}  \sum_{u,v \in \mathcal{V}_{sub}} 
     \big( \sigma(m_{uv})\log \sigma(m_{uv}) \\
      & + (1-\sigma(m_{uv}))\log (1-\sigma(m_{uv}))
     \big),
    \end{aligned}
\end{align}
where the hyper-parameter $\eta$ is the confidence size controlling how many entries in $m_{uv}$ could be free of penalization.

\begin{algorithm}[htb]
\caption{\textit{CAMA-subgraph} for structure attack.} 
\label{alg:structure-mask}
\KwIn{Graph $G = (\bm{A}, \bm{X})$ with $n$ nodes; the ground truth label $c_{gt}$ of graph $G$; 
ranked nodes vector of the predicted label $\bm{U}^c$;
subgraph proportion $p\%$; 
victim model $f_{\bm{\Theta}}$; 
total training epoch number $T$;
the perturbation budget $\Delta$;
}
\KwOut{Modified Adjacency matrix $\bm{A}'$.}

Initialize perturbation candidate subgraph $\mathcal{V}_{sub}=\{u|u \in U^c[:n_{sub}]\}$, where $n_{sub}=\lfloor p\%|\mathcal{V}| \rfloor$.

\For{$t$ in $1,2,...,T$}{
    // Train subgraph mask
    
    $\min_{\bm{M}} l_{atk}= l_{cw}+\lambda_{ent}l_{ent}$;
    
    // Generate the adversarial example
    
    select top $\Delta$ perturbations $\bm{M}_{\Delta} \sim Bernoulli(\bm{M})$;
    
    \For{$(u,v) \in \{(u,v)|m_{uv} \in \bm{M}_{\Delta}$\}}{
        $a'_{uv} \leftarrow 1-a_{uv}$;
    }
    
    $G' \leftarrow (\bm{A'}, \bm{X})$;
    
    // Test the adversarial example
    
    \If{$\argmax_c f_{\bm{\Theta}}(G') \neq c_{gt}$}
    {
    break;
    }
    
}

\textbf{return} $\bm{A}'$;
\end{algorithm}

Algorithm~\ref{alg:structure-mask} shows the whole attacking process of structure attack with subgraph mask training, and we denote it as \textbf{\textit{CAMA-subgraph}}.
First, we select top-ranked nodes in $\bm{U}^c$ to formulate a subgraph and limit the edge perturbation within the subgraph in line 1.
Secondly, for each training epoch, we minimize the attack loss $l_{atk}$ to train the subgraph mask $\bm{M}$ as shown in line 4.
Then, we select the top-ranked mask $\bm{M}_{\Delta}$ within the perturbation budget $\Delta$ in line 6. In lines 7-9, we flip edges for nodes pair selected in $\bm{M}_{\Delta}$ to generate the adversarial example.
Finally, we test the attack performance of generated adversarial examples in lines 11-12.

\subsection{Complexity Analysis}\label{sec:complexity}
We analyze the complexity of the proposed framework by using \textit{CAMA} as an example. Given a graph with $n$ nodes as target, the main complexity lies in the preparation of inputs:
\begin{itemize}
\item The original nodes ranking matrix $\bm{U}_{CAM}^{orig}$ (\textbf{Algorithm~\ref{alg:rank}}): The complexity of line 1 is $\mathcal{O} (Cn\log(n)) = \mathcal{O} (n\log(n))$, since the number of classes is always much less than that of nodes. Then the complexity from line 2 to 6 is $\mathcal{O}(Cn)$. Thus the total complexity of Algorithm~\ref{alg:rank} is $\mathcal{O} (n\log(n) + Cn)$;
\item Feature noise $\epsilon_j$, where $j=1,2,...,D_L$: The complexity of getting all $\epsilon$ is $\mathcal{O} (nD_L + nK) = \mathcal{O} (nD_L)$, since $K$ is seleted from $D_L$;
\item Similarity matrix $\bm{S}$: The complexity of having similarity matrix is $\mathcal{O} ( n^{2}D_L )$.
\end{itemize}

Then we analyze the complexity of Algorithm~\ref{alg:feature} and Algorithm~\ref{alg:structure} accordingly, note that all constraints have no effects on the complexity since they can be checked in constant time:

\xhdr{Feature attack (Algorithm~\ref{alg:feature})}
The complexity from line 3 to line 5 is $\mathcal{O} (r)$. Thus, the total complexity of Algorithm~\ref{alg:feature} is combining it with $\bm{U}^c$ and all $\epsilon$, which is $\mathcal{O} (n \times \max(D_L, \log(n)))$.

\xhdr{Structure attack (Algorithm~\ref{alg:structure})}
The complexity from line 2 to line 11 is $\mathcal{O} (\min(n^{2}, \Delta))$. Thus, combining with the complexity of similarity matrix $\bm{S}$, the total complexity of Algorithm~\ref{alg:structure} is $\mathcal{O} (\min(n^{2}D_L, \Delta)) = \mathcal{O} (\Delta)$, since the modification budget $\Delta$ is controlled to restrict the access from attackers and strictly smaller than $n^{2}$.

Through our analysis of the complexity above, we can find that \textit{CAMA} enjoys computational efficiency, especially in comparison with the complexity of target GNNs.

\section{Experiments}
In this section, 
we evaluate the effectiveness of our proposed methods on the graph classification task under the white-box and black-box settings.
We further conduct sensitivity analysis for hyper-parameters and provide the poisoning black-box attack performance.

\subsection{Experimental Setups}
\xhdr{Datasets}
We evaluate our attack strategies on five chemical graph classification benchmarks: MUTAG, PROTEINS, NCI1, COX2~\cite{Morris+2020}, NCI-H23, and three social network datasets: IMDB-BINARY, IMDB-MULTI, DBLP\_v1. Among chemical graphs, node features consist of node attributes and node labels: 
in PROTEINS and COX2, we use both node labels and attributes, while in the others, we only use one-hot node labels as node features. 
For social networks, node features are initialized with the node degree. 
The dataset statistics can be found in Table \ref{tab:Dataset statistics}.

\begin{table}[htbp]
\renewcommand{\arraystretch}{1.2}
\caption{Dataset statistics.} \label{tab:Dataset statistics}
\vspace{-4mm}
\begin{center}
\resizebox{0.97\columnwidth}{!}{%
\begin{tabular}{l c c c c  }
\toprule[1pt]
\textbf{Dataset}  & \textbf{\#Graphs}   &  \textbf{\#Classes}  & \textbf{Avg. \#Nodes} &   \textbf{Avg. \#Edges} \\
\midrule
MUTAG     & 188      & 2  & 17.93& 19.79\\
PROTEINS &   1,113  &2 &	39.06	&	72.82 \\	
NCI1    &   4,110   &2  &29.87&	32.3\\
COX2  & 467  & 2 & 41.22&43.45\\
NCI-H23 & 40,353	& 2	& 26.07	& 28.10 \\
\midrule
IMDB-BINARY & 1,000 & 2 & 19.77 &96.53  \\
IMDB-MULTI  & 1,500 &3 &13.00&  65.94 \\
DBLP\_v1 & 19,456 & 2& 10.48 & 19.65 \\
\bottomrule
\end{tabular}
}
\end{center}
\vspace{-4mm}
\end{table}

\begin{table*}[!t]
\centering
\caption{Summary of the change in classification accuracy (in \%) compared to the clean graph under \textbf{white-box attack} for \textbf{chemical datasets}. Lower is better.
Best performances are shown in \textbf{bold} markers.
\label{tab:whit-box attack result}}
\resizebox{\textwidth}{!}{%
\begin{tabular}{ l c c c c c c c c c c c c  c c c c}
\toprule
    \textbf{Dataset} & \multicolumn{4}{c}{\textbf{MUTAG}} & \multicolumn{4}{c}{\textbf{PROTEINS}} & \multicolumn{4}{c}{\textbf{NCI1}}  & \multicolumn{4}{c}{\textbf{COX2}} \\
\cmidrule(lr){2-5}\cmidrule(l){6-9}\cmidrule(l){10-13}\cmidrule(l){14-17}
Models & GCN & GIN-0 & IGNN & g-U-Nets & GCN & GIN-0 & IGNN & g-U-Nets & GCN & GIN-0 & IGNN & g-U-Nets& GCN & GIN-0 & IGNN & g-U-Nets\\
Clean &    83.04 & 89.85  & 81.46 & 88.89	& 78.17 & 77.81  & 77.99& 77.54 & 78.98	& 77.59 & 75.06 & 72.24 & 88.87 & 83.51 & 83.51 & 83.08  \\
\midrule
\multicolumn{17}{c}{\textbf{\textit{Feature Attack}}} \\
\textit{Random}\cite{dai2018adversarial} & -5.35 & -5.29 & -7.43 & -7.02 & -1.26 & -0.63 & -4.04 & -0.90&-16.06 & -19.03 & -33.92 & -55.57 & -17.32 & -3.42 & -7.05 & -11.54 \\ 
\textit{Degree}\cite{tong2012gelling} & -4.82 & -7.40 & -7.43 & -8.66 & -1.61 & -0.81 & -4.58 & -0.99 &  -17.27 & -23.63 & -37.40 & -59.37 & -22.67 & -4.28 & -8.54 & -13.90\\ 
\textit{PageRank}~\cite{ma2020towards} & -4.30 & -3.18 & -6.37 & -6.99 & -1.53 & -0.72 & -4.67 & -0.99 & -18.96 & -21.99 & -37.83 & -59.95 & -25.48 & -6.00 & -10.28 & -12.40 \\
\textit{Betweenness}~\cite{ma2020towards} & -5.88 & -8.97 & -6.90 & -8.13 & -1.44 & -0.72 & -4.49 & -0.90 & -15.70 & -19.17 & -34.87 & -57.64 & -17.10 & -4.08 & -7.70 & -13.91 \\
\textit{GC-RWCS}~\cite{ma2020towards} & -6.43 & -7.92 & -6.90 & -8.13 & -1.53 & -0.81 & -4.13 & -0.99 & -16.64 & -22.58 & -33.72 & -57.40 & -20.10 & -3.88 & -7.91 & -13.70 \\
\textit{RWCS}~\cite{ma2020towards} & -5.35 & -7.92 & -6.90 & -7.57 & -1.71 & -0.63 & -4.58 & -0.99 & -17.52 & -24.21 & -35.47 & -58.74 & -23.75 & -5.99 & -8.78 & -13.06 \\
\midrule 
\textit{\textbf{CAMA}}&-10.64 &\textbf{-9.53} &\textbf{-10.12} &\textbf{-11.78} & -2.24 & -1.44 &\textbf{ -6.56 }& \textbf{-2.25}&\textbf{-33.58} & \textbf{-36.08 }& -56.74 & -69.61 &\textbf{-52.68} & -9.69 & \textbf{-27.83}& \textbf{-27.64} \\ 
\textit{\textbf{CAMA-Grad}}&\textbf{-11.70} & \textbf{-9.53} &\textbf{-10.12}& -11.73 & \textbf{-2.60} & \textbf{-1.53} & -6.29 & \textbf{-2.25}&-31.70 & -35.57 &\textbf{-56.76}&\textbf{-69.90}& -52.23 & \textbf{-15.47} & -22.89 & -24.40 \\
\midrule
\midrule
\multicolumn{17}{c}{\textbf{\textit{Structure Attack}}}
\\
\textit{Random}~\cite{dai2018adversarial}& -4.82 & -16.43 & -5.26 & -2.13 & -0.99 & -4.13 & -1.53 & -0.54 & -9.49 & -10.97 & -6.37 & -4.31 & -6.43 & -3.84 & -2.14 & -4.93  \\ 
\textit{Degree}~\cite{tong2012gelling}& 8.48 & -16.43 & -7.92 & -3.27 & -0.72 & -6.91 & -1.53 & -0.09 & -8.08 & -15.13 & -5.79 & -4.31 & -6.87 & -9.83 & -4.07 & -5.56  \\ 
\textit{GradArgmax}\cite{dai2018adversarial}& -7.98 & -43.33 & -7.37 & -2.13 & -1.88 & -7.63 & -2.96 & -1.08 & -10.90 & -12.31 & -10.85 & -7.45 & -17.17 & -16.24 & -13.08 & -11.99 \\
\textit{PR-BCD}~\cite{PR-BCD} & -17.54 & -55.76 & \textbf{-19.68} & -6.99 & -4.85 & -33.25 & -3.42 & -3.78 & -47.84 & -19.85 & -46.50 & -23.36 & -55.22 & -52.55 & -30.80 & -32.11 \\
\midrule 
 \textbf{\textit{CAMA}} & -11.08 & -47.07 & -11.64 & -9.18 & -3.23 & -9.44 & -2.88 & -1.80 & -20.68 & -22.43 & -15.74 & -9.88 & -22.48 & -18.89 & -13.93 & -12.64 \\
\textbf{\textit{CAMA}-Grad} & -11.64 & -50.20 & -12.72 & -5.85 & -2.78 & -9.16 & -3.24 & -1.53 & -23.51 & -22.29 & -16.69 & -8.76 & -24.86 & -18.85 & -13.28 & -15.82 \\
\textbf{\textit{CAMA-subgraph}} & \textbf{-25.44} & -74.44 & -18.62 & -7.49 & \textbf{-6.91} & \textbf{-33.43} & \textbf{-6.02} & -3.69 & \textbf{-61.44} & -54.40 & \textbf{-49.68} & \textbf{-23.77} & \textbf{-57.85} & \textbf{-58.43} & \textbf{-34.88} & -33.82 \\
\textbf{\textit{CAMA-subgraph-Grad}} & -23.86 & \textbf{-75.55} & \textbf{-19.68} & \textbf{-10.24} & -5.84 & -32.81 & -5.93 & \textbf{-3.87} & -61.24 & \textbf{-55.67} & -48.98 & -21.92 & -54.41 & \textbf{-58.43} & -32.29 & \textbf{-35.32} \\
\bottomrule
\end{tabular}
}
\end{table*}

\begin{table*}[!t]
\centering
\caption{Summary of the change in classification accuracy (in \%) compared to the clean graph under \textbf{white-box attack for \textbf{social networks}}. Lower is better. 
Best performances are shown in \textbf{bold} markers.
\label{tab:social network attack result}}
\begin{tabular}{lccccccccc}
\toprule
\textbf{Dataset} & \multicolumn{3}{c}{\textbf{IMDB-BINARY}} & \multicolumn{3}{c}{\textbf{IMDB-MULTI}} & \multicolumn{3}{c}{\textbf{DBLP\_v1}} \\                       
\cmidrule(lr){2-4}\cmidrule(lr){5-7}\cmidrule(lr){8-10}
Models  & \multicolumn{1}{c}{GCN} & \multicolumn{1}{c}{GIN} & \multicolumn{1}{c}{g-U-Nets} & \multicolumn{1}{c}{GCN} & \multicolumn{1}{c}{GIN} & \multicolumn{1}{c}{g-U-Nets} & \multicolumn{1}{c}{GCN} & \multicolumn{1}{c}{GIN} & \multicolumn{1}{c}{g-U-Nets} \\
Clean & 73.67 & 74.22 & 73.89 & 50.00 & 50.59& 48.30 & 90.47 & 91.52 & 93.55\\
\midrule
\textit{Random}~\cite{dai2018adversarial}                  & -0.78                    & -7.55                                        & -0.78                     & -0.96                    & -9.48                               & -0.45 &-0.37 & -3.21 & -0.32 \\
\textit{Degree}~\cite{tong2012gelling}                  & -1.67                    & -18.00          & -2.78                     & -1.63                    & -14.37                   & -2.37&-0.37 & -4.02 & -0.31\\
\textit{GradArgmax}~\cite{dai2018adversarial}              & -4.34                    & -19.22                   & -3.33                     & -2.82                    & -14.52                   & -1.11 & -0.32 & -4.29 & -0.51\\
\textit{PGD}~\cite{dai2018adversarial}                      & -2.57                    & -22.82                   & -1.79                     & -2.00                    & -29.12                   & -0.57 & -1.33 & -11.24 & -1.10   \\
\textit{PR-BCD}~\cite{PR-BCD}& -5.87 & -17.42 & -7.99 & -4.80 & -20.79 & -2.50 & -0.90 & -8.40 & -0.78 \\
\textit{ReWatt}~\cite{10.1145/3447548.3467416} & -6.07 & -3.22  & -6.99  & -6.53 & -2.79  & -3.03  & -1.50 &-2.18  & \textbf{-2.08}\\
\midrule
\midrule
\textbf{\textit{CAMA}} & -2.11 & -15.22 & -2.33 & -3.11 & -11.48 & -1.19 & -0.72 & -4.42 & -0.50 \\
\textbf{\textit{CAMA-Grad}} & -2.78 & -15.55 & -1.56 & -3.26 & -13.33 & -1.04 & -0.80 & -4.98 & -0.60 \\
\textbf{\textit{CAMA-subgraph}} & \textbf{-7.77} & -21.52 & -8.59 & -7.73 & \textbf{-30.39} & \textbf{-4.10} & -1.44 & \textbf{-11.54} & -1.18 \\
\textbf{\textit{CAMA-subgraph-Grad}} & -7.37 & \textbf{-22.82} & \textbf{-8.79} & \textbf{-8.07} & -30.19 & -3.97 & \textbf{-1.52} & -10.89 & -1.13 \\      
\bottomrule
\end{tabular}
\end{table*}

\begin{table*}[!t]
\centering
\caption{Summary of the change in classification accuracy (in \%) compared to the clean graph under \textbf{black-box attack}. Lower is better.
\label{tab:black-box attack result}
}
\resizebox{1\textwidth}{!}{%
\begin{tabular}{ l c c c c c c c c c c c c}
\toprule
    \textbf{Dataset} & \multicolumn{3}{c}{\textbf{MUTAG}} & \multicolumn{3}{c}{\textbf{PROTEINS}} & \multicolumn{3}{c}{\textbf{NCI1}}  & \multicolumn{3}{c}{\textbf{COX2}} \\
\cmidrule(lr){2-4}\cmidrule(l){5-7}\cmidrule(l){8-10}\cmidrule(l){11-13}
Models  & GIN-0 & IGNN & g-U-Nets& GIN-0 & IGNN & g-U-Nets & GIN-0 & IGNN & g-U-Nets & GIN-0 & IGNN & g-U-Nets\\
Clean & 89.85  & 81.46 & 88.89 & 77.81  & 77.99& 77.54	& 77.59 & 75.06 & 72.24 & 83.51 & 83.51 & 83.08 \\
\midrule
\multicolumn{13}{c}{\textit{\textbf{
Feature Attack}}} \\
\textit{Random}~\cite{dai2018adversarial}  & -2.13 &{-4.24} & -4.85 & -0.54 & -3.59 & -0.45  & -6.59 & -9.32 & -13.14  & -2.56 & -7.26 & -9.19 \\ 
\textit{Degree}~\cite{tong2012gelling}   & -2.66 &{-4.24} &\textbf{-6.52}  & -0.63 & -4.13 & -0.54 & -9.46 & -10.19 & -14.89  & -3.85 & -7.69 & -10.69 \\ 
\textit{PageRank}~\cite{ma2020towards} & -2.66 & -3.71 & -4.85 & -0.45 & -3.96 & -0.45 & -9.10 & -12.82 & -15.26 & -5.57 & -8.33 & -10.06 \\
\textit{Betweenness}~\cite{ma2020towards} & -3.71 & -3.71 & \textbf{-6.52} & -0.45 & -3.87 & -0.36 & -7.57 & -12.14 & -13.97 & -3.64 & -6.83 & -11.55 \\
\textit{GC-RWCS }~\cite{ma2020towards}& -2.66 & -3.71 & \textbf{-6.52} & -0.54 & -3.60 & -0.45 & -7.47 & -11.51 & -13.82 & -3.88 & -6.84 & -9.83 \\
\textit{RWCS}~\cite{ma2020towards} & -2.66 & -3.71 & -5.97 & -0.45 & -3.96 & -0.45 & -9.29 & -11.36 & -14.16 & -5.99 & -6.84 & -10.27 \\
\textit{\textbf{CAMA}}&\textbf{-4.24}& {-6.93}  & -5.38 & -0.90 & -5.57 & \textbf{-1.26}  & \textbf{-17.13} & \textbf{-23.72} & \textbf{-24.28} & \textbf{-12.88} & \textbf{-22.04}& \textbf{-22.26}\\ 
\textit{\textbf{CAMA-Grad}} &-3.71 & \textbf{-8.01} & -4.27  &\textbf{-0.99}& \textbf{-5.84} & -1.17  & -15.23 & -22.65 & -24.06  & -12.44 & -19.03 & -20.35 \\ 
\midrule
\midrule
\multicolumn{13}{c}{\textit{\textbf{Structure Attack}}} \\
\textit{Random}~\cite{dai2018adversarial}    & -16.43 & -5.26 & -2.13 & -4.13 & -1.53 & -0.54 & -10.97 & -6.37 & -4.31  & -3.84 & -2.14 & -4.93\\ 
 \textit{Degree}~\cite{tong2012gelling} & -16.43 & -7.92 & -3.27  & -6.91 & -1.53 & -0.09  & -15.13 & -5.79 & -4.31  & -9.83 & -4.07 & -5.56 \\ 
\textit{GradArgmax}~\cite{dai2018adversarial}  & -12.75 & -9.53 & -2.72  & -5.48 & -1.17 & -0.90  & -8.88 & -6.67 & -3.75 & -9.82 & -4.48 & -5.12 \\ 
\textit{ReWatt}~\cite{10.1145/3447548.3467416} & -6.84 & -3.68 & -9.12 & -2.61 & -0.81 & \textbf{-1.17} & -7.57 & -4.94 & -8.23 & -7.70 & -2.57 & -13.07 \\
\textit{Grabnel}~\cite{wan2021adversarial} & -42.39 & -11.11 & -2.66 & -8.53 & -2.25 & -0.90 & -26.32 & -15.55 & -5.52 & -11.13 & -6.41 & -12.24 \\
\textit{\textbf{CAMA}} & -47.07 & -11.64 & \textbf{-9.18} & -9.44 & -2.88 & -1.35 & -22.43 & -15.74 & \textbf{-9.88} & -18.89 & \textbf{-13.93} & -12.64 \\
\textit{\textbf{CAMA-Grad}} & -50.20 & \textbf{-12.72} & -5.85 & -9.16 & \textbf{-3.24} & -1.08 & -22.29 & -16.69 & -8.76 & -18.85 & -13.28 & \textbf{-15.82} \\
\textit{\textbf{CAMA-subgraph}} & -59.53 & -11.69 & -3.77 & \textbf{-24.89} & -2.25 & -0.99 & -25.84 & -15.81 & -7.52 & \textbf{-53.73} & -12.20 & -11.57 \\
\textbf{\textit{CAMA-subgraph-Grad}} & \textbf{-60.03} & -10.64 & -5.44 & -23.63 & -2.61 & \textbf{-1.17} & \textbf{-26.69} & \textbf{-17.42} & -9.32 & -53.51 & -13.68 & -8.56 \\
\bottomrule
\end{tabular}
}
\end{table*}

\xhdr{Graph Classifiers} 
We use four state-of-the-art GNNs for graph classification: GCN, GIN, IGNN, and g-U-Nets.
Only one fully connected layer is adopted for all configurations, and no dropout layer is used after graph pooling. The same global sum-pooling readout function is applied for all models.
For GCN, we use 5 GCN convolutional layers. For GIN, we set $\epsilon=0$ (also called GIN-0) and use 5 GIN convolution layers. For IGNN, we use 3 IGNN convolution layers and tune hyper-parameter $\kappa \in \{0.7, 0.98\}$. We fix the size of hidden dimensions as 64.
g-U-Nets have a different architecture due to their hierarchical nature. Here, we use the node representation of the last layer before the readout function to calculate the CAM heat-map matrix. We apply four (graph pooling) gPool layers with 90\%, 70\%, 60\%, and 50\% node proportions and ignore the max-pooling layer in its readout function since global max-pooling is poorer at localization compared to GAP~\cite{zhou2016learning}. We implement these GNNs with Pytorch Geometric (PyG)\footnote{https://github.com/rusty1s/pytorch\_geometric}.

\xhdr{Baselines}
We compare our methods with representative feature attack baselines which select perturbation nodes from various perspectives (\textit{Random}, \textit{Degree}, \textit{Betweenness}, \textit{RWCS}, etc.).
For all baselines under feature attacks, the same feature noises in Section~\ref{Feature Attack} are added to selected nodes, the difference only lies in the nodes selected process.
We also compare \textit{CAMA} with representative white-box (\textit{PGD}, \textit{PR-BCD}) and black-box (\textit{ReWatt}, \textit{Grabnel}) structure attack methods.
Every baseline we compared either released source code or made it available upon request.
The detailed baselines are described as follows.
\begin{itemize}
    \item (structure/feature) \textit{Random} \cite{dai2018adversarial}: \textit{Random} randomly selects nodes to perturb and edges to insert/delete.
    \item (structure/feature) \textit{Degree} \cite{tong2012gelling}: \textit{Degree} chooses nodes with top degrees and insert/delete edges among them.
    \item (structure) \textit{GradArgmax} \cite{dai2018adversarial}: \textit{GradArgmax} greedily selects perturbation edges by gradients of each pair of nodes, which works only for structure attack. 
    \item (structure) \textit{PGD}~\cite{pgd}: \textit{PGD} performs project gradient descent topology attacks and is an effective white-box attack algorithm.
    \item (structure) \textit{PR-BCD}~\cite{PR-BCD}: \textit{PR-BCD} conducts sparsity-aware first-order optimization attacks based on randomized block coordinate descent and is able to attack larger graphs.
    \item (structure) \textit{ReWatt}~\cite{10.1145/3447548.3467416}: \textit{ReWatt} conducts rewiring operations to perform structure attacks and uses reinforcement learning to find the optimal rewiring operations. We select \textit{ReWatt} as the representative of the state-of-the-art black-box optimization baseline.
    \item (structure) \textit{GRABNEL}~\cite{wan2021adversarial}: \textit{GRABNEL} is a powerful black-box attack method on graph classification tasks based on the bayesian optimization.
    \item (feature) \textit{PageRank}~\cite{ma2020towards}: \textit{PageRank} is a graph centrality metric. Here, we attack nodes with the top-ranked Pagerank scores.
    \item (feature) \textit{Betweenness}~\cite{ma2020towards}: \textit{Betweenness} is a graph centrality metric. Here, we attack nodes with the top-ranked Betweenness scores.
    \item (feature) \textit{RWCS}~\cite{ma2020towards}:
    \textit{RWCS} is a practical feature attack algorithm based on an importance score similar to PageRank by using the connection between the GNNs' backward propagation and random walks.
    \item (feature) \textit{GC-RWCS}~\cite{ma2020towards}: 
     \textit{GC-RWCS} is a variant of \textit{RWCS}, which uses the greedy correction procedure on top of the RWCS strategy.
     \item (feature, structure) \textit{GraphAttacker}~\cite{chen2021graphattacker}:
    GraphAttacker performs attacks based on the generative adversarial network and three key components: the multi-strategy attack generator, the similarity discriminator, and the attack discriminator.
     \item (structure) \textit{Projective Ranking}~\cite{10.1145/3459637.3482161}:
     \textit{Projective Ranking} exploits mutual information to consider the long-term benefits of perturbations and generates adversarial samples.
     \item (feature, structure) \textit{Attack on the HGP Model}~\cite{tang2020adversarial}:  \textit{Attack on the HGP Model} aims to fool the pooling operator in hierarchical GNN-based graph classification models.
    
\end{itemize}

\xhdr{Perturbation restrictions and hyper-parameters}
For feature attack, we set feature adjusted magnitude $\lambda=0.1$. We select $10\% $ of nodes in one graph to perturb, and $10\%$ of features are modified for each dataset. 
For structure attack, we set the perturbation budget $\Delta=\lceil  10\% |E_i|\rceil  $ for each graph $G_i$, where $|E_i|$ denotes the number of edges in graph $G_i$. For \textit{ReWatt}, the number of rewiring operations is set to $ \lfloor 0.5 \Delta \rfloor$ with at least one rewiring, which is kept the same setting as~\cite{10.1145/3447548.3467416}.
Besides, in the similarity restriction, we use the first hidden layer to calculate nodes similarity $\bm{h}^{emb}=\vh^{(1)}$, fix $s_2=0.95$  and tune $s_1\in\{0.95,0.9,1\}$.
For \textit{CAMA-subgraph}, we set total training epochs as 30, the subgraph graph proportion $p\% = 50\%$, $\lambda_{ent}=1$.

We conduct the untargeted attack and evaluate them on test graphs. Specifically, we perform 10-fold cross-validation in each classification process and report the average validation accuracy within the cross-validation. This configuration follows \cite{xu2019how} on graph classification, resulting from the unstable training of small-sized datasets such as MUTAG.

\subsection{Adversarial Attack on Graph Classification}
We first compare \textit{CAMA} and \textit{CAMA-subgraph} to multiple baselines under the white-box attack. We train on clean graphs for each graph classifier, generate perturbed graphs on validation sets, and calculate prediction accuracy using the trained graph classifiers. 
Full results under the white-box setting for chemical datasets are provided in Table \ref{tab:whit-box attack result}, for social networks are demonstrated in Table~\ref{tab:social network attack result}.
Additional experimental results on the NCI-H23 dataset are shown in Table~\ref{tab: NCI-H23 white}.

In feature attack, our proposed methods perform better by a high margin on all datasets and all graph classification models, which implies our methods can select the most influential nodes for graph classification tasks.
In structure attack, \textit{CAMA} and \textit{CAMA-subgraph} outperform the other baselines in all situations.
Meanwhile, the subgraph mask training algorithm (\textit{CAMA-subgraph}) outperforms the simple heuristic flip edge method (\textit{CAMA}) by a large margin.
Actually, the choice of \textit{CAMA} and \textit{CAMA-subgraph} is to balance the attack efficiency and effectiveness.
These results demonstrate the high attack effectiveness of \textit{CAMA}.
More interestingly, the grad version \textit{CAMA-Grad} achieves excellent performance close to \textit{CAMA} but does not guarantee better performance.

We also observe that the attack results vary from different datasets and graph classifiers. The accuracy decreases the least on the PROTEINS dataset when suffering attacks. Interestingly, graph classifiers tend to behave differently when they are attacked by structure and feature perturbations. For example, IGNN is more robust facing structural perturbations while more vulnerable under feature attack.

\begin{table}[htbp]
\centering
\caption{White-box attack results on NCI-H23.}
\label{tab: NCI-H23 white}
\resizebox{0.98\columnwidth}{!}{
\begin{tabular}{lcccc}
\toprule
Models & GCN & GIN & IGNN & GNet \\
Clean & 95.28 & 94.94 & 95.10 & 94.99 \\
\midrule
\multicolumn{5}{c}{\textbf{\textit{Feature Attack}}} \\
\textit{Random}~\cite{dai2018adversarial} & -1.97 & -0.10 & -4.02 & -0.86 \\
\textit{Degree}~\cite{tong2012gelling} & -2.09 & -0.12 & -5.33 & -1.08 \\
\textit{PageRank}~\cite{ma2020towards} & -2.25 & -0.09 & -3.42 & -1.16 \\
\textit{Betweenness}~\cite{ma2020towards} & -1.90 & -0.16 & -3.17 & -0.93 \\
\textit{GC-RWCS}~\cite{ma2020towards} & -2.04 & -0.13 & -3.03 & -0.91 \\
\textit{RWCS}~\cite{ma2020towards} & -2.01 & -0.12 & -3.23 & -0.99 \\
\textit{CAMA} & -4.35 & \textbf{-1.02} & -12.80 & \textbf{-5.88} \\
\textit{CAMA-Grad} & \textbf{-5.77} & -0.79 & \textbf{-12.92} & -4.89 \\
\midrule
\multicolumn{5}{c}{\textbf{\textit{Structure Attack}}} \\
\textit{Random}~\cite{dai2018adversarial} & -0.75 & -0.09 & -0.20 & -0.07 \\
\textit{Degree}~\cite{tong2012gelling} & -0.58 & -0.08 & -0.17 & -0.04 \\
\textit{GradArgmax}~\cite{dai2018adversarial} & -0.77 & -0.05 & -0.34 & -0.10 \\
\textit{PR-BCD}~\cite{PR-BCD}& -4.69 & \textbf{-8.55} & -1.50 & -0.09 \\
\textit{CAMA} & -1.17 & -0.83 & -0.57 & -0.10 \\
\textit{CAMA-Grad} & -1.57 & -0.08 & -0.50 & -0.13 \\
\textit{CAMA-subgraph} & \textbf{-14.36} & -1.43 & \textbf{-1.91} & -0.13 \\
\textit{CAMA-subgraph-Grad} & -14.31 & -0.98 & -1.64 & \textbf{-0.21} \\
\bottomrule
\end{tabular}
}
\end{table}

\begin{table}[htbp]
\centering
\caption{Black-box feature attack results on NCI-H23.}
\label{tab: NCI-H23 black}
\begin{tabular}{lccc}
\toprule
Models & GIN & IGNN & GNet \\
Clean & 94.94 & 95.10 & 94.99 \\
\midrule
\textit{Random}~\cite{dai2018adversarial} & -0.02 & -1.24 & -0.11 \\
\textit{Degree}~\cite{tong2012gelling} & -0.03 & -1.72 & -0.11 \\
\textit{PageRank}~\cite{ma2020towards} & -0.03 & -0.76 & -0.13 \\
\textit{Betweenness}~\cite{ma2020towards} & -0.04 & -0.95 & -0.12 \\
\textit{GC-RWCS}~\cite{ma2020towards} & -0.02 & -0.69 & -0.12 \\
\textit{RWCS~\cite{ma2020towards}} & -0.03 & -0.62 & -0.12 \\
\textit{CAMA} & -0.05 & -4.00 & -0.20 \\
\textit{CAMA-Grad} & \textbf{-0.07} & \textbf{-4.32} & \textbf{-0.22} \\
\bottomrule
\end{tabular}
\end{table}

\begin{table}[htbp]
\centering
\caption{Black-box structure attack results on NCI-H23.}
\label{tab: NCI-H23 black struc}
\begin{tabular}{lccc}
\toprule
Models & GIN & IGNN & GNet \\
Clean & 94.94 & 95.10 & 94.99 \\
\midrule
\textit{Random}~\cite{dai2018adversarial} & -0.00 & -0.20 & -0.03 \\
\textit{Degree}~\cite{tong2012gelling} & -0.09 & -0.17 & -0.04 \\
\textit{Gradargmax}~\cite{dai2018adversarial} & -0.03 & -0.19 & -0.06 \\
\textit{ReWatt}~\cite{10.1145/3447548.3467416} & -0.00 & -0.00 & 0.00 \\
\textit{Grabnel}~\cite{wan2021adversarial} & 0.05 & -0.22 & - \\
\textit{CAMA} & -0.07 & -0.27 & -0.07 \\
\textit{CAMA-Grad} & -0.09 & -0.52 &\textbf{-0.09} \\
\textit{CAMA-subgraph} & \textbf{-0.23} & -0.52 & -0.07 \\
\textit{CAMA-subgraph-Grad} & -0.21 & \textbf{-0.58} & -0.08 \\
\bottomrule
\end{tabular}
\end{table}

\subsection{Transferability of Attack}
In real-world applications, model parameters usually are not available. 
Thus, to evaluate \textit{CAMA} under a more realistic and general situation and further explore the transferability of various attacking methods, we validate our attack strategies under the black-box attack setting for four datasets.
Specifically, we use GCN as the surrogate model, generate adversarial examples by targeting GCN, and then evaluate the other GNNs on the perturbed graphs. The detailed results are provided in Table~\ref{tab:black-box attack result}. Additional experimental results on the NCI-H23 dataset are shown in Table~\ref{tab: NCI-H23 black} (feature attack) and Table~\ref{tab: NCI-H23 black struc} (structure attack).

First, we could see our approaches surpass the other baselines in most situations. The perturbations generated by \textit{CAMA} and \textit{CAMA-subgraph} consistently demonstrate strong transferability on four graph classification datasets under the black-box attack setting. 
For \textit{CAMA-subgraph}, we could also see a significant performance improvement of \textit{CAMA-subgraph} over \textit{CAMA}. The calculation of the ranked CAM matrix and the selected subgraph is important. As a result, the attack performance of a black-box attack may exceed a white-box attack due to an efficient ranked CAM matrix of the GCN surrogate model.
Moreover, we could see that the attack performance of \textit{ReWatt} is unstable. It does work with some datasets, like NCI1, while it fails for the other datasets. 
Second, compared with the white-box attack, our approaches have a more significant advantage over baselines like \textit{GradArgmax}. This indicates that our methods have a more vital attack ability when transferring to other GNNs.
Besides, the results show that perturbations against a surrogate model with typical architecture could also generalize to the hierarchical graph classifier like g-U-Nets.

\subsection{Localization Effectiveness}

\begin{table}[htbp]
\caption{Comparison of average degrees for selected nodes under various feature attack methods.}
\centering
\vspace{-2mm}
\label{tab: avg degree cama}
\resizebox{0.99\columnwidth}{!}{
\begin{tabular}{lccccc}
\toprule
Dataset & \textit{Random} & \textit{RWCS} &\textit{GC\_RWCS} & \textit{\textbf{CAMA}} & \textit{\textbf{CAMA-Grad}} \\
\midrule
MUTAG & 2.23 &2.99 & 2.72& 2.12 & 2.28 \\
COX2 & 2.00 & 3.18 & 2.49& 2.26 & 2.35 \\
PROTEINS & 3.74 & 4.73 &4.13 &	4.11 & 3.96  \\
NCI1 & 2.14 & 2.96 &2.64 &	1.92 & 1.91  \\
\bottomrule
\vspace{-3mm}
\end{tabular}
}
\end{table}

\begin{figure}[htbp]
\centering
\includegraphics[width=0.99\columnwidth]{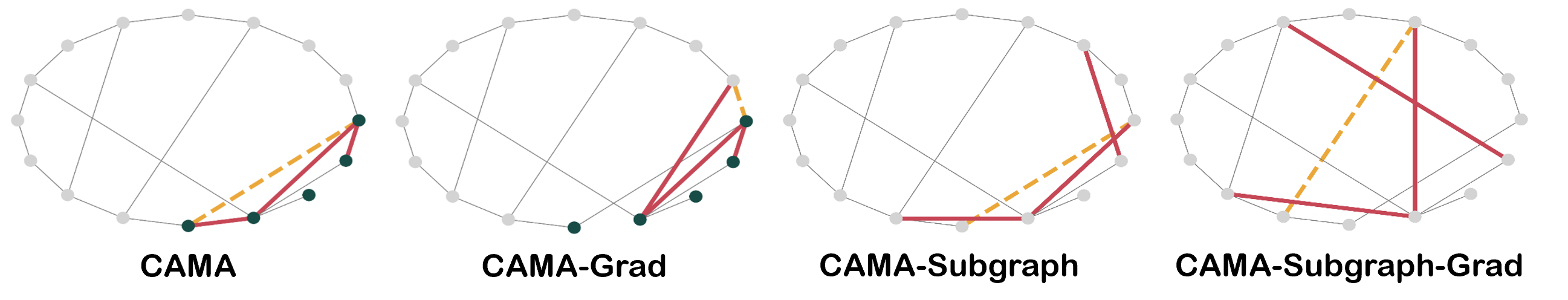}
\caption{\textbf{An example of \textit{CAMA} on MUTAG dataset with edge attack proportion=$20\%$.}
Green nodes are selected by \textit{CAMA}, indicating their strong influences on graph classification. 
Added edges are shown in red lines and deleted edges are shown in orange dashed lines.
}
\label{fig: comparison CAMA}
\end{figure}

Here, we show the effectiveness of \textit{CAMA} in the global-to-local attack challenge raised in Sec.~\ref{sec: challenge}.
For the feature attack, we show the average node degree selected by \textit{CAMA} and other baselines in Table~\ref{tab: avg degree cama}.
In comparison to baselines such as \textit{RWCS}, the average degree of nodes selected by \textit{CAMA} and \textit{CAMA\_Grad} are lower and closer to that of \textit{Random}, which indicates our proposed methods are more unnoticeable.
What's more, we visualize the perturbation nodes and edges selected by \textit{CAMA} in Figure~\ref{fig: comparison CAMA} as a comparison with Figure~\ref{fig: comparison}.
The edges chosen by \textit{CAMA} are not trapped in one node compared to \textit{Degree}, \textit{GradArgmax}, and \textit{ReWatt}.

\begin{figure}[htbp!]
\centering
\includegraphics[width=0.7\columnwidth]{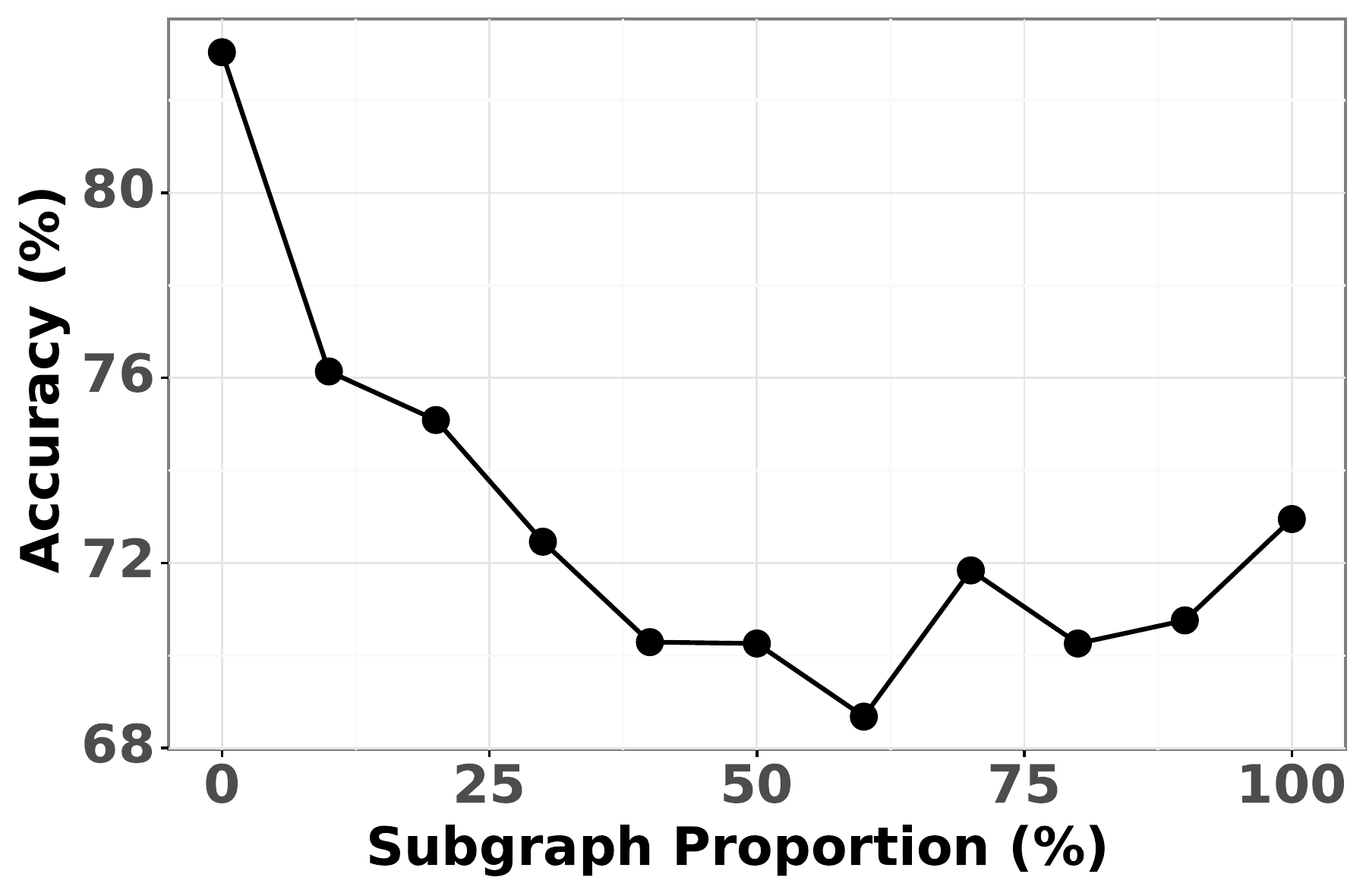}
	\captionof{figure}{Line-plot for attack performance under \textit{CAMA-subgraph} for different subgraph proportion. We record 10-fold testing results on the MUTAG dataset using GCN as the graph classifier. Lower is better. \label{fig:subgraph}}
\vspace{-2mm}
\end{figure}

\xhdr{Insights of target nodes chosen by \textit{CAMA}}
We compare the top 5 nodes selected by \textit{CAMA} and \textit{Degree} and report statistics in Table~\ref{tab:insight}.
The relatively small average degree and closeness centrality value differentiate \textit{CAMA} from centrality-based methods.
Through the total variation and number of edges, we find that nodes chosen by \textit{CAMA} have higher connectivity and smoothness (smaller total variation).
Besides, we provide an example of edge perturbations on baselines in Figure 3.

\begin{table}[htbp]
\caption{Statistics for selected nodes by \textit{CAMA} and \textit{Degree}.}
\label{tab:insight}
\centering
\resizebox{\columnwidth}{!}{
\begin{tabular}{lcccc}
\toprule
Method & Avg. Degree & Avg. Closeness & Total Variation & No. Edges \\
\midrule
\textit{Degree} & 2.8 & 0.25 & 12 & 1 \\
\textit{\textbf{CAMA}} & 2.4 & 0.22 & 8 & 2 \\
\bottomrule
\end{tabular}
}
\vspace{-3mm}
\end{table}

\subsection{Sensitivity Analysis\label{ablation}}

\begin{figure*}[thb!]
\centering
\includegraphics[width=0.3\linewidth]{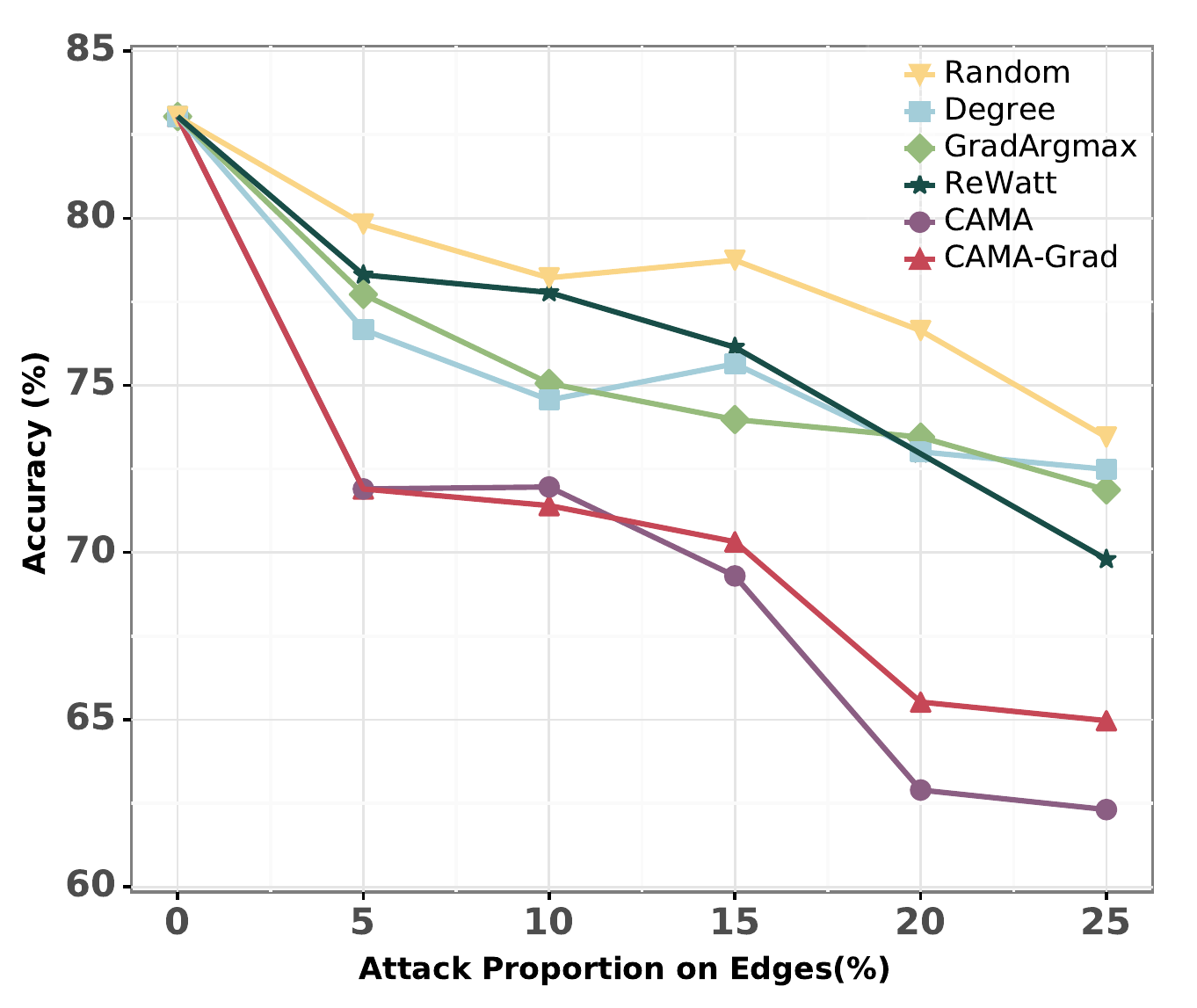}
\includegraphics[width=0.3\linewidth]{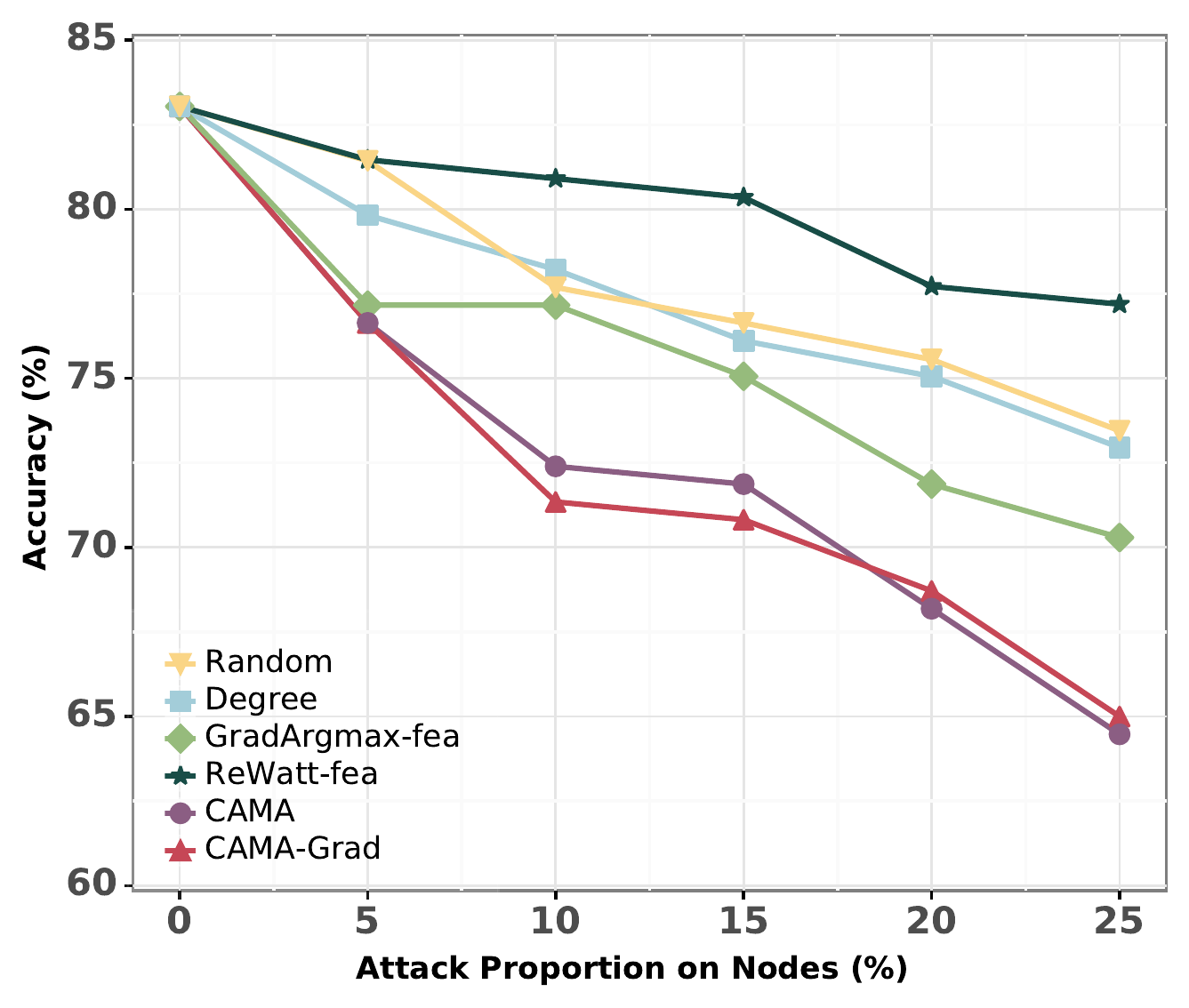}
\includegraphics[width=0.375\linewidth]{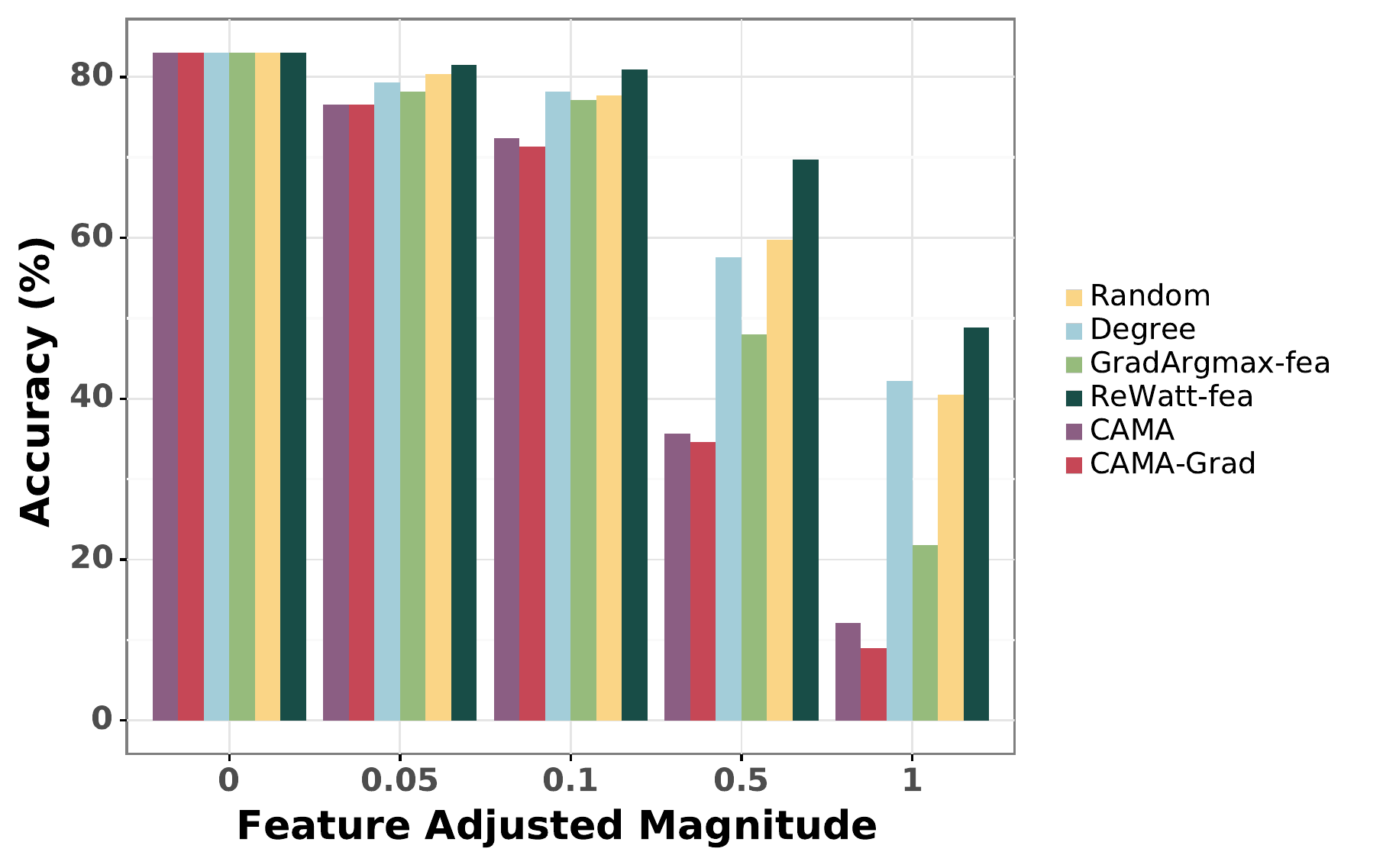}
\vspace{-3mm}
\caption{\textbf{Attack results with different perturbation hyper-parameters.} All experiments are conducted on the MUTAG dataset using GCN. Lower accuracy is better. Left: attack results with increasing perturbation proportion of edges. Middle: attack results with increasing perturbation proportion of nodes. Right: attack results with increasing adjusted magnitude values.}
\label{fig: ablation}
\vspace{-3mm}
\end{figure*}

\xhdr{Sensitivity analysis for subgraph proportion $p$ in \textit{CAMA-subgraph}}
The choice of subgraph proportion in \textit{CAMA-subgraph} is crucial. A larger proportion means more perturbation candidates but also more noise, while a smaller proportion may face perturbation candidates deficiency.
An efficient subgraph selection could help the attacker localize the essential subgraph nodes and edges.
Figure~\ref{fig:subgraph} shows the attack performance of \textit{CAMA-subgraph} with various subgraph proportions on MUTAG. 
We could see a clear drop tendency when the subgraph proportion gets smaller from 100\%, which indicates the effectiveness of locating the subgraph with the ranked CAM vector. 
For MUTAG, the best proportion is 60\%, and the accuracy drop is 14.36\% under this structure attack perturbation setting.

\xhdr{Sensitivity Analysis for Hyper-parameter $s_1$ and $s_2$ in \textit{CAMA}}
We perform a sensitivity analysis over $s_1$ and $s_2$ in Table~\ref{tab:sensitivity} and set GIN as the victim model on the MUTAG dataset. 
$s_1$ controls the edge insertion and $s_2$ controls the edge deletion. $s_1=1, s_2=0$ represents no restriction on edge insertion/deletion.
We could find that controlling the edge insertion is more helpful for successful attacks in contrast to edge deletion.

\begin{table}[htb!]
\renewcommand{\arraystretch}{1.11}
\caption{Sensitivity analysis for hyper-parameter $s_1$ and $s_2$. Lower is better.} \label{tab:sensitivity}
\vspace{-2mm}
\begin{center}
\resizebox{1\columnwidth}{!}{%
\begin{tabular}{cccccccc}
\hline
Hyper-parameter & clean & 0 & 0.2 & 0.4 & 0.6 & 0.8 & 1 \\ \hline
$s_1$ (fix $s_2$=0) & 83.04 & 67.72 & 67.72 & 66.64 & 67.16 & 66.64 & 71.40 \\
$s_2$ (fix $s_1$=1) & 83.04 & 71.40 & 71.40 & 71.93 & 71.93 & 71.93 & 72.46 \\ \hline
\end{tabular}
}
\end{center}
\end{table}

\xhdr{Perturbations budget for white-box attack}
We analyze the changes in accuracy with respect to the perturbation budget $\Delta$ and the adjusted magnitude $\lambda$ in Figure~\ref{fig: ablation}. Not surprisingly, the prediction accuracy decreases with a higher number of perturbations or larger values of adjusted magnitude.
In all settings of hyper-parameters, we can observe that \textit{CAMA} and \textit{CAMA-Grad} show remarkable advantages over all the other baselines. 
Meanwhile, from the figure on the right, we can observe the accuracy drops dramatically when the adjusted magnitude $\lambda$ gets larger for \textit{CAMA} and \textit{CAMA-Grad}.

\begin{figure}[htbp]
\centering
\includegraphics[width=0.8\columnwidth]{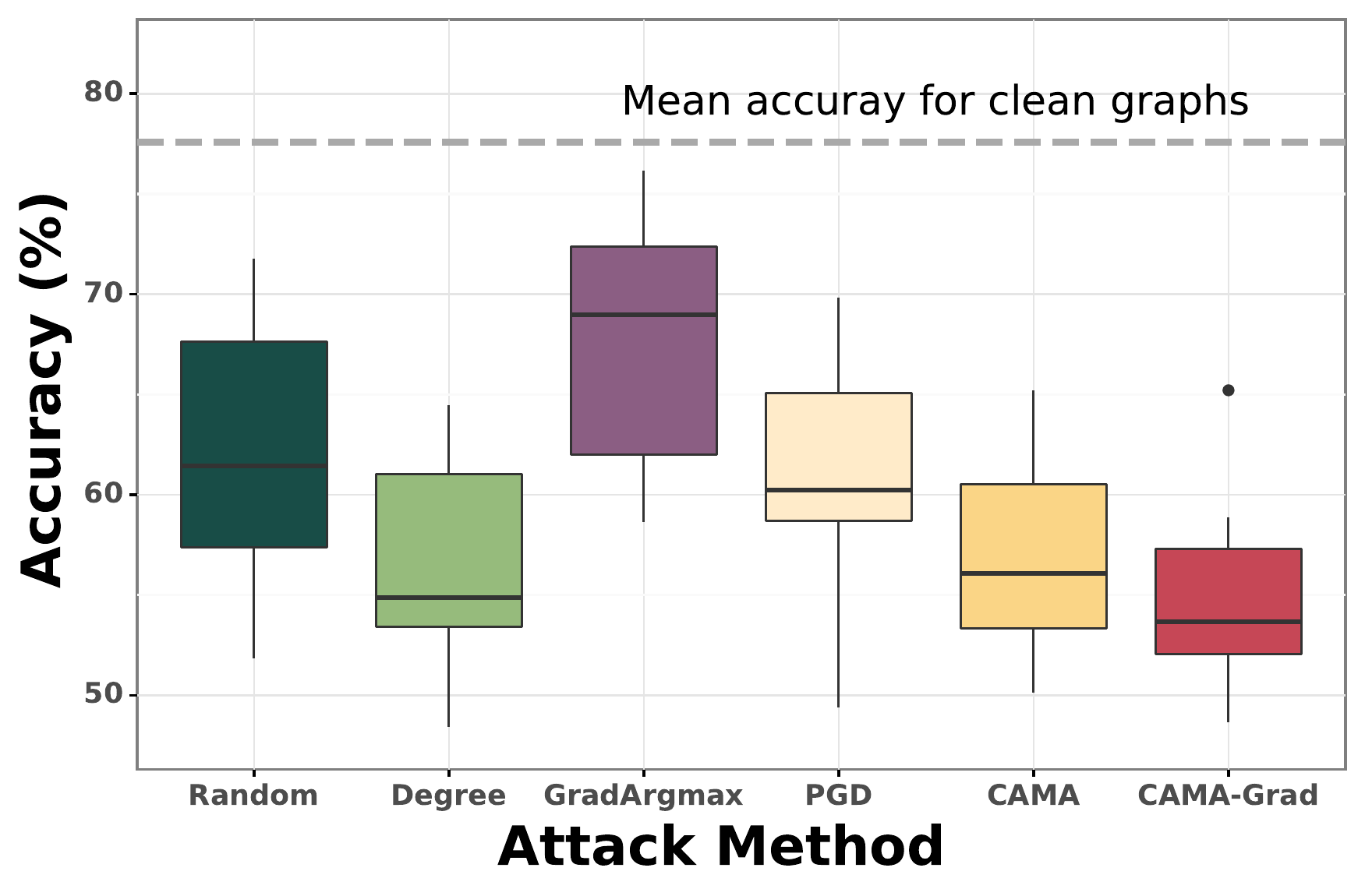}
    \vspace{-2mm}
	\captionof{figure}{Box-plot for \textbf{poisoning} structural perturbations under black-box attack. We use GIN as the victim model and record 10-fold testing results on the NCI1 dataset. Lower is better. \label{fig:poison}}
\vspace{-4mm}
\end{figure}

\subsection{Poisoning Black-box Attack}
We also evaluate our methods under poisoning black-box attacks.
We select GIN as the victim model and retrain it on perturbed graphs generated from the surrogate GCN.
Additionally, we compare \textit{CAMA} with a more powerful attacker, project gradient descent topology attack (\textit{PGD})~\cite{pgd}.  \textit{PGD} was originally designed for node classification tasks. We extend its application domain to graph classification.
We use cross entropy loss and fix epoch numbers to 10 in our experiment under \textit{PGD} topology attack.
Figure~\ref{fig:poison} shows the final attack results.
Coordinating with the results of the evasion attack above, the strong transferability of \textit{CAMA} and \textit{CAMA-Grad} still concludes. However, the method using purely gradient information like \textit{GradArgmax} and \textit{PGD} may damage the attacking performance when transferring to other models.

\subsection{Computational Efficiency Analysis}

To cooperate with our complexity analysis in Section~\ref{sec:complexity}, we demonstrate the computational efficiency of \textit{CAMA} and \textit{CAMA-subgraph} in Table~\ref{tab:timeCost} by reporting the average running time over 10 times in comparison with all baselines. 
We can find that \textit{CAMA} can finish within 1 seconds, which is consistent with our complexity analysis and implies that the scalability of our proposed approaches could not be an issue.
The time cost of \textit{CAMA-subgraph} is comparable with \textit{ReWatt}. Both methods need end-to-end training, but \textit{CAMA-subgraph} has better attack performance.

\begin{table}[htbp]
\renewcommand{\arraystretch}{1.11}
\caption{Running time ($s$) comparison overall baseline methods on MUTAG using GCN. We report the $10$ times average running time within 10-fold cross-validation.\label{tab:timeCost}}
\begin{center}
\resizebox{0.99\linewidth}{!}{%
\begin{tabular}{ c c c c c c c}
\hline
Models  & \textit{Random} & \textit{Degree} & \textit{GradArgmax} & \textit{ReWatt} & \textit{\textbf{CAMA}} & \textit{\textbf{CAMA-Subgraph}} \\
\hline
Structure    &0.3694      & 0.3820  & 0.3429 &30.0120 & 0.7888& 34.2288 \\
Feature    &0.6252     & 0.6110  & 0.7538 & - &0.6747 & - \\
\hline
\end{tabular}
}
\end{center}
\vspace{-6mm}
\end{table}

\section{Conclusion}
We revisit adversarial attacks on GNNs for graph classification in this paper. We establish a general attack framework focusing on graph classification which considers comprehensive attack settings under white-box and black-box attacks and performs both structure and feature attacks. 
We first estimate the importance of nodes towards the graph classification by Class Activation Mapping and its variant. Then, we heuristically design algorithms to generate adversarial examples for both feature and structure attacks with the ranking information of nodes.
Experiments show that the proposed attack strategies significantly outperform existing approaches on various graph classifiers under multiple settings. 
Our general framework can also serve as a simple yet novel baseline for future works in evaluating the robustness of graph classification tasks.




%

\ifCLASSOPTIONcaptionsoff
  \newpage
\fi




\printbibliography 

%



%


\begin{IEEEbiography}[{\includegraphics[width=1in,height=1.25in,clip,keepaspectratio]{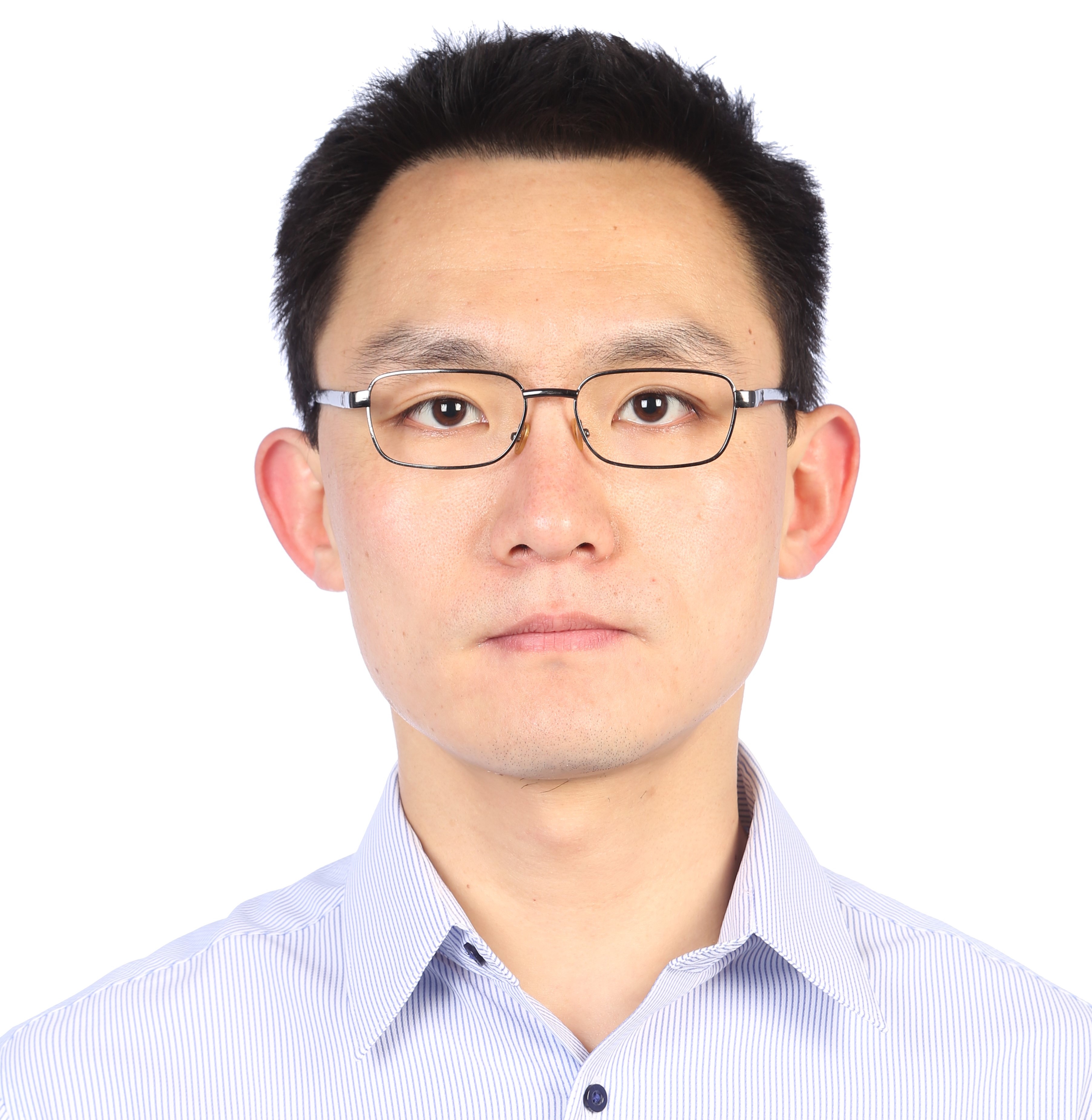}}]{Xin Wang} is currently an Assistant Professor at the Department of Computer Science and Technology, Tsinghua University. He got both his Ph.D. and B.E degrees in Computer Science and Technology from Zhejiang University, China. He also holds a Ph.D. degree in Computing Science from Simon Fraser University, Canada. His research interests include relational media big data analysis, multimedia intelligence and recommendation in social media. He has published over 100 high-quality research papers in top journals and conferences including IEEE TPAMI, IEEE TKDE, ACM TOIS, ICML, NeurIPS, ACM KDD, ACM Web Conference, ACM SIGIR and ACM Multimedia etc. He is the recipient of 2017 China Postdoctoral innovative talents supporting program. He received the ACM China Rising Star Award in 2020 and IEEE TCMC Rising Star Award in 2022.
\end{IEEEbiography}

\begin{IEEEbiography}[{\includegraphics[width=1in,height=1.25in,clip,keepaspectratio]{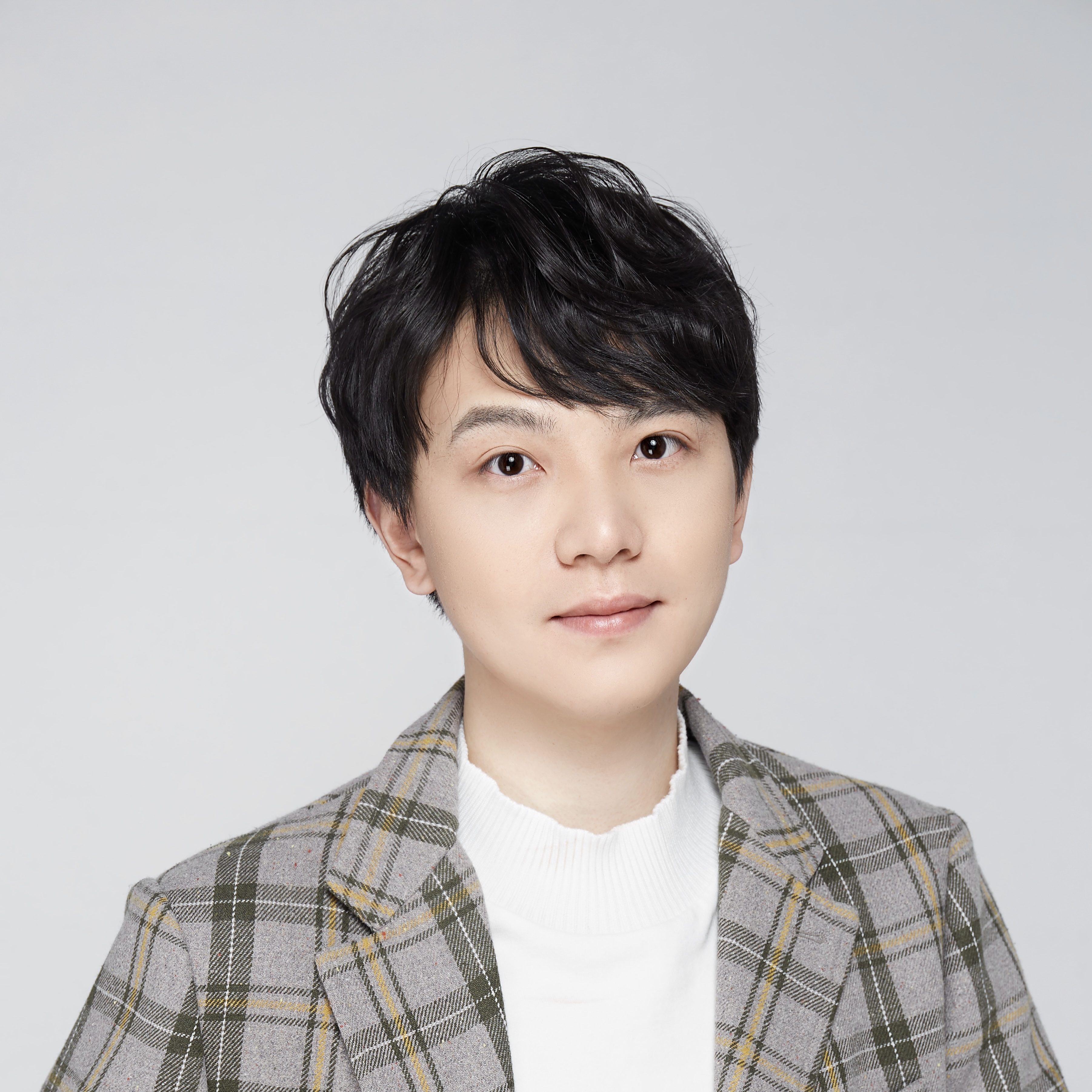}}]{Heng Chang} is currently pursuing a Ph.D. Degree in the Tsinghua-Berkeley Shenzhen Institute at Tsinghua University. He received his B.S. from the Department of Electronic Engineering, Tsinghua University in 2017. His research interests focus on representation learning, adversarial robustness, and machine learning on graph/relational structured data. He has published several papers in prestigious conferences/journals including NeurIPS, AAAI, TheWebConf, TKDE, TPAMI, \etc. 
\end{IEEEbiography}

\begin{IEEEbiography}[{\includegraphics[width=1in,height=1.25in,clip,keepaspectratio]{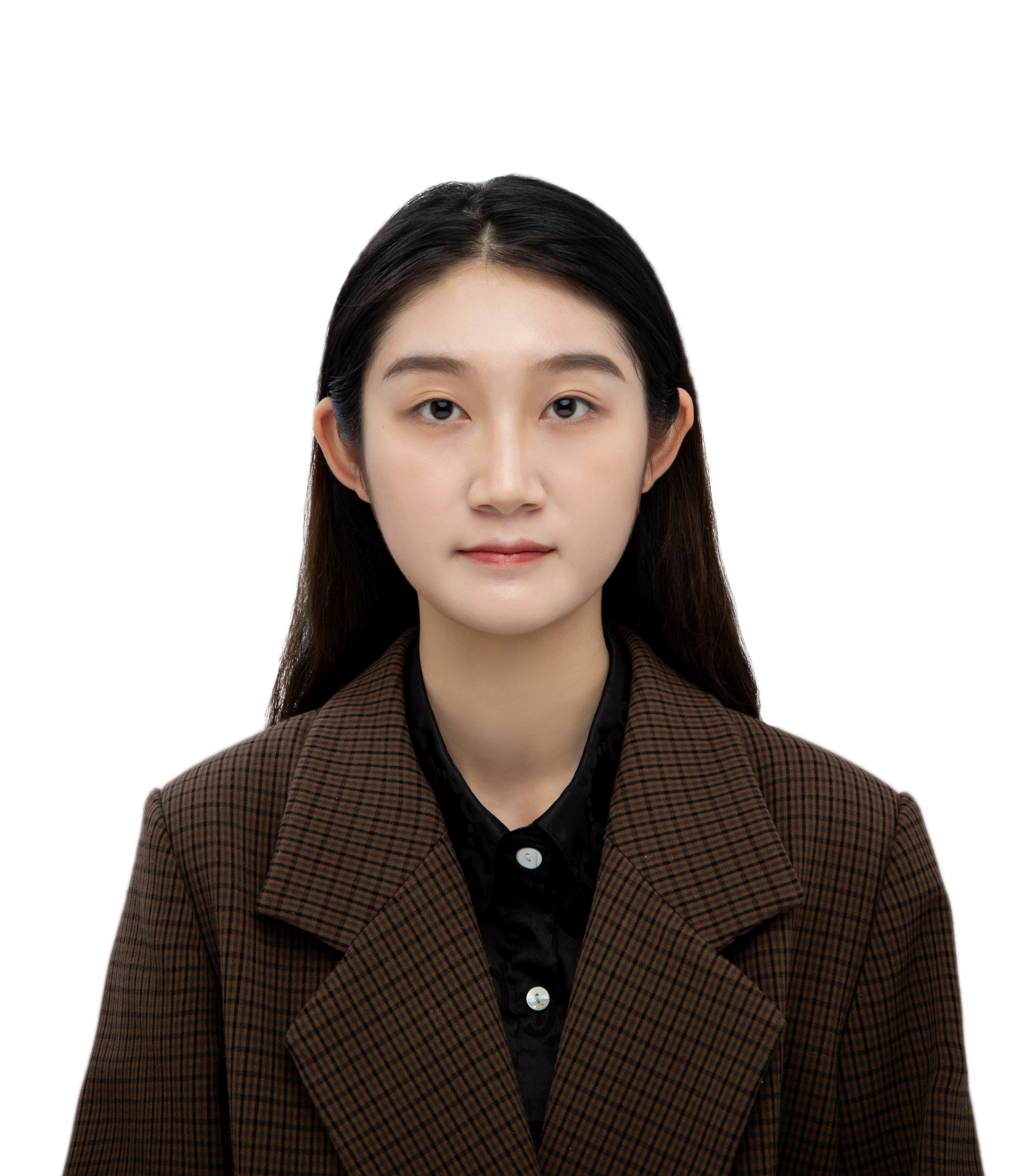}}]{Beini Xie}  is currently an M.A. student in the Tsinghua-Berkeley Shenzhen Institute at Tsinghua University. She received her B.S from the Statistical Department, Renmin University in 2020. Her research interests include adversarial robustness, machine learning and neural architecture search on graph/relational structured data.
 \vspace{-10mm}
\end{IEEEbiography}

\begin{IEEEbiography}[{\includegraphics[width=1in,height=1.25in,clip,keepaspectratio]{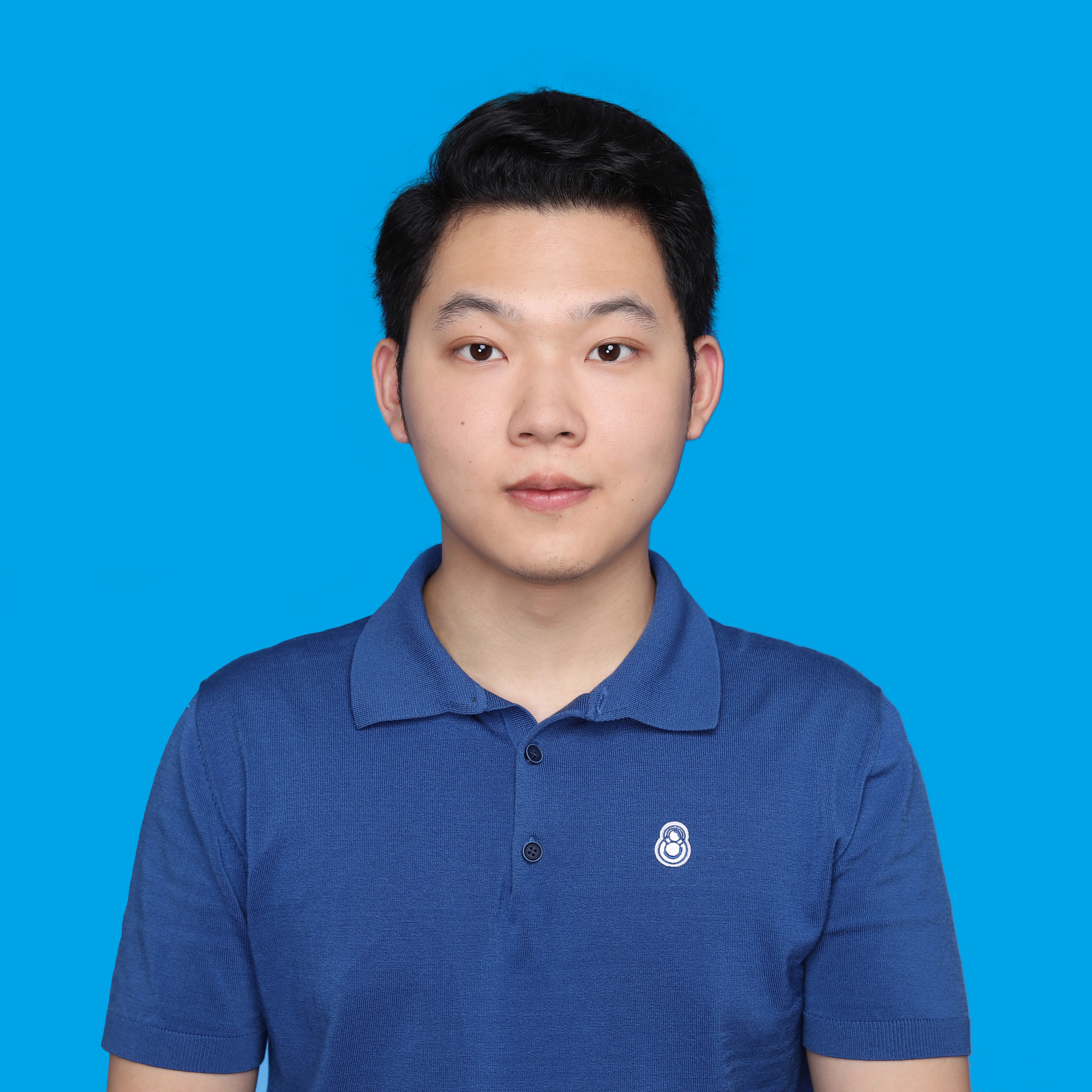}}]{Tian Bian} is currently a Ph.D. student at The Chinese University of Hong Kong. He received his B.E. degree from Southwest University in 2018 and his Master's degree from Tsinghua University in 2021. His research interests focus on graph neural networks and their applications such as social media analysis. He has served as a reviewer for ICML, NeurIPS, AAAI, etc.
\end{IEEEbiography}

\begin{IEEEbiography}[{\includegraphics[width=1in,height=1.25in,clip,keepaspectratio]{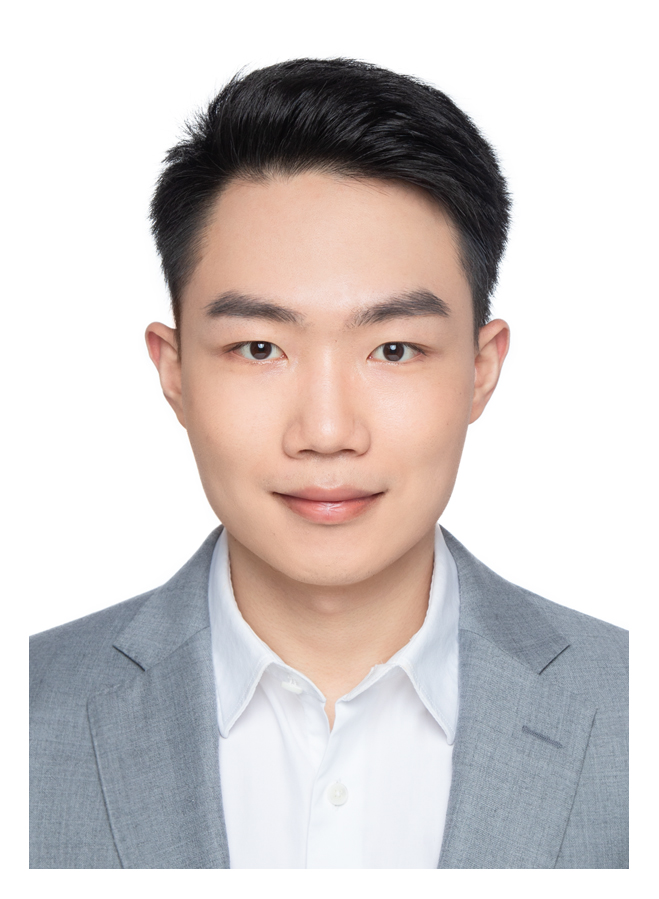}}]{Shiji Zhou}
is currently a Ph.D. candidate at Tsinghua University. He received his B.S. from the Chinese Academy of Science-Beihang Hua Luogeng Mathematics Honors Class, Beihang University in 2017. His research covers online learning and multi-objective optimization, as reflected in his publications on top-tier conferences and journals, including NeurIPS, AISTATS, TMM. He has been the co-organizer of the 3rd Human in the Loop Learning (HILL) workshop on ICML. He has served as a reviewer for ICML, NeurIPS, JSAC, etc. 
\end{IEEEbiography}

\begin{IEEEbiography}[{\includegraphics[width=1in,height=1.25in,clip,keepaspectratio]{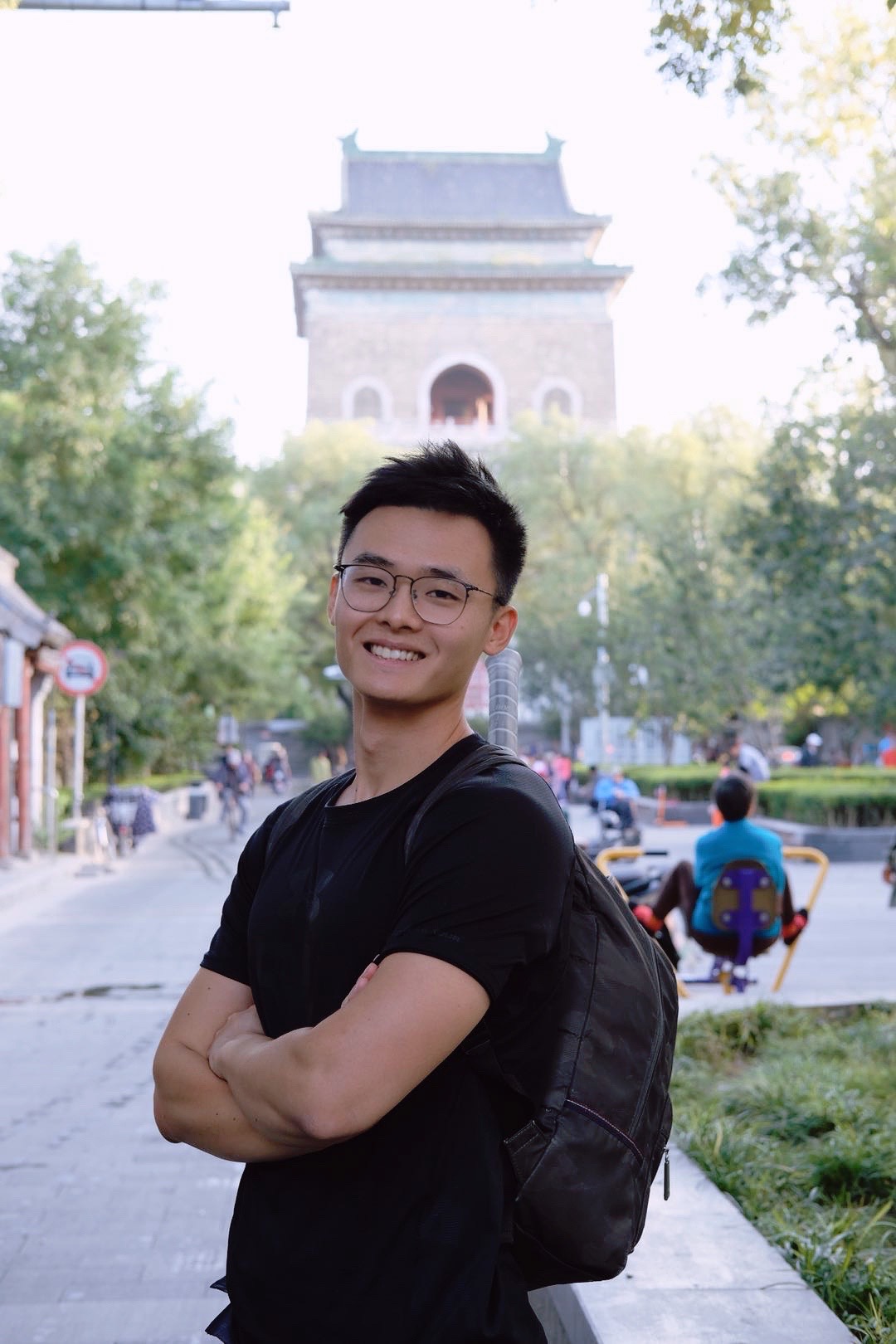}}]{Daixin Wang}
received his Ph.D. degree in computer science and technology from Tsinghua University, Beijing, China, in 2018, and he is currently working as an algorithm expert on Ant Group. He has authored or coauthored more than 10 papers in conferences such as KDD, AAAI, and IJCAI, and journals such as the IEEE Transactions on Multimedia. His research interests include graph learning and multimodal learning.
\vspace{-10mm}
\end{IEEEbiography}

\begin{IEEEbiography}[{\includegraphics[width=1in,height=1.25in,clip,keepaspectratio]{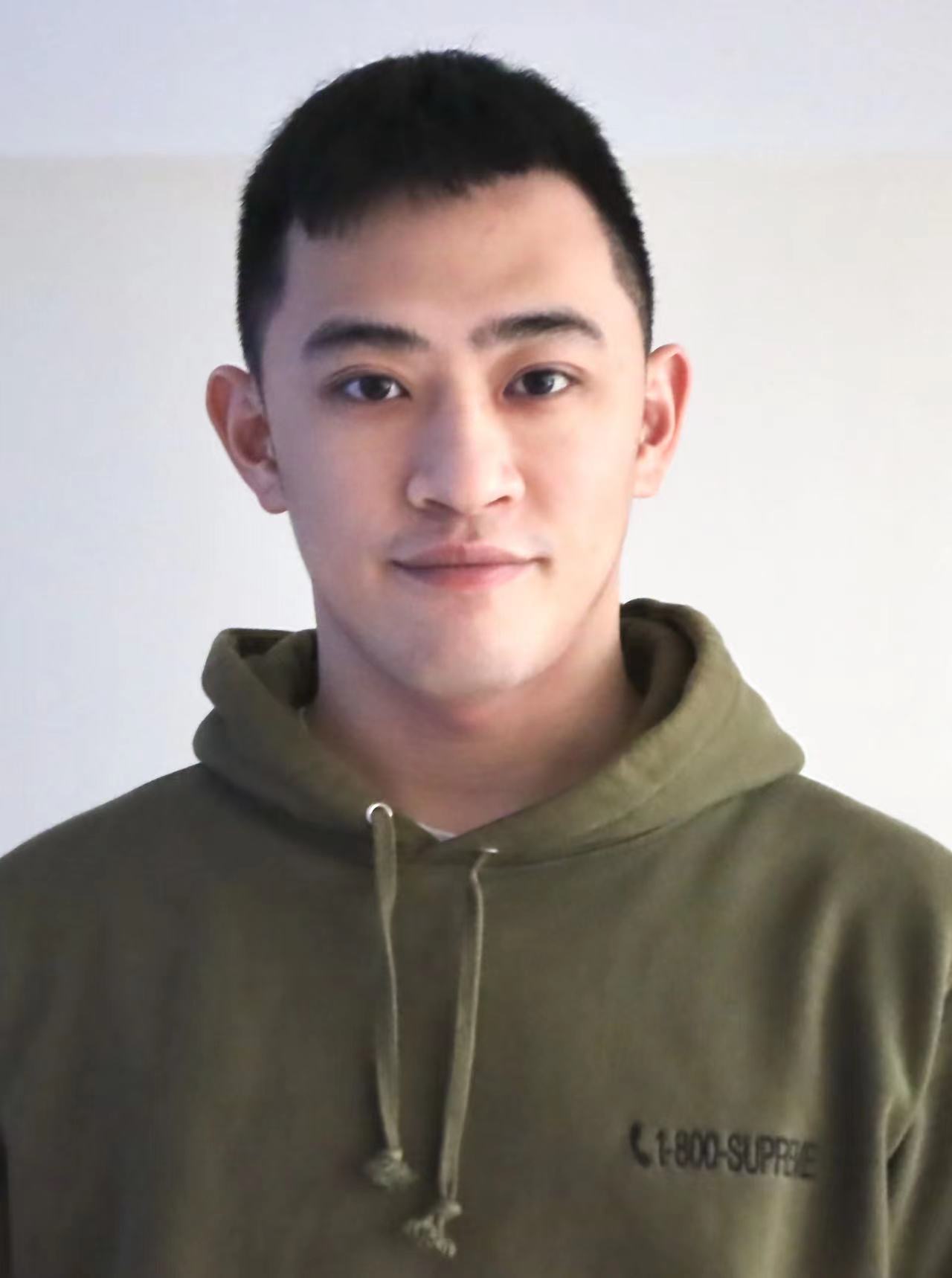}}]{Zhiqiang Zhang}
is currently a Staff Engineer at Ant Group. His research interests mainly focus on graph machine learning. He has led a team to build an industrial graph machine learning system, AGL, in Ant Group. He has published more than 30 papers in top-tier machine learning and data mining conferences, including NeurIPS, VLDB, SIGKDD, and AAAI.
\end{IEEEbiography}

\begin{IEEEbiography}[{\includegraphics[width=1in,height=1.25in,clip,keepaspectratio]{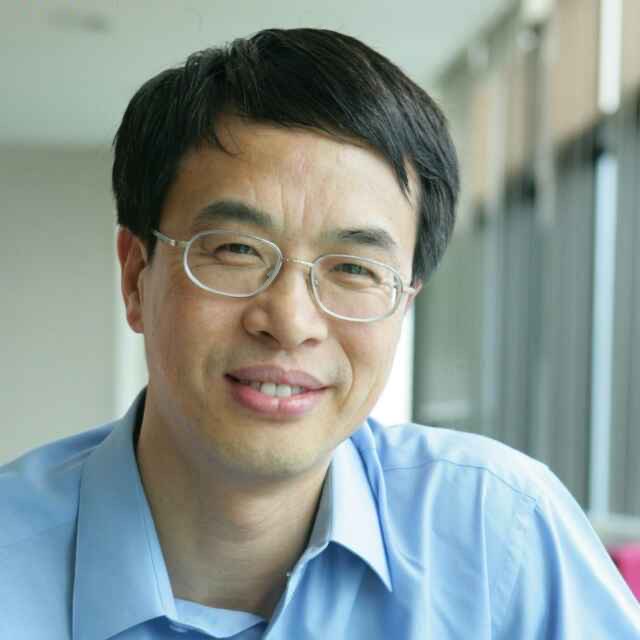}}]{Wenwu Zhu} is currently a Professor in the Department of Computer Science and Technology at Tsinghua University, the Vice Dean of National Research Center for Information Science and Technology. Prior to his current post, he was a Senior Researcher and Research Manager at Microsoft Research Asia. He was the Chief Scientist and Director at Intel Research China from 2004 to 2008. He worked at Bell Labs New Jersey as a Member of Technical Staff during 1996-1999. He received his Ph.D. degree from New York University in 1996.

His current research interests are in the area of data-driven multimedia networking and multimedia intelligence. He has published over 350 referred papers and is inventor or co-inventor of over 50 patents. He received ten Best Paper Awards, including ACM Multimedia 2012 and IEEE Transactions on Circuits and Systems for Video Technology in 2001 and 2019.

He served as EiC for IEEE Transactions on Multimedia (2017-2019), the chair of the steering committee for IEEE Transactions on Multimedia (2019-2021), and the Associate EiC for IEEE Transactions for Circuits and Systems for Video Technology. He serves as General Co-Chair for ACM Multimedia 2018 and ACM CIKM 2019, respectively. He is an ACM Fellow, AAAS Fellow, IEEE Fellow, SPIE Fellow, and a member of The Academy of Europe (Academia Europaea).
\end{IEEEbiography}








\end{document}